\documentclass[aps,preprint,
superscriptaddress,
showpacs,preprintnumbers,amsmath,amssymb,nofootinbib,longbibliography]{revtex4-1}
\usepackage[
colorlinks=true, 
pdfstartview=FitV, 
linkcolor=blue, 
citecolor=magenta, 
urlcolor=blue, 
bookmarks=true,
bookmarksnumbered=true,
pdftitle={},
pdfauthor={Masaru Hongo (RIKEN)}
]{hyperref}
\usepackage[dvips]{graphicx}
\usepackage{bm,latexsym,amsmath,amssymb,amsfonts,mathrsfs,textcomp}
\usepackage{color}
\input{colordvi.tex}

\newcommand{\diag}{\mathop{\mathrm{diag}}}

\newcommand\Lcal{\mathcal{L}}
\newcommand\Ncal{\mathcal{N}}
\newcommand\Scal{\mathcal{S}}
\newcommand\Hcal{\mathcal{H}}

\newcommand\Dcal{\mathcal{D}}
\newcommand\Acal{\mathcal{A}}
\newcommand\Ecal{\mathcal{E}}
\newcommand\eff{\mathrm{eff}}

\newcommand{\SO}{\mathrm{SO}}
\newcommand{\U}{\mathrm{U}}
\renewcommand{\O}{\mathrm{O}}

\newcommand{\so}{\mathrm{so}}

\newcommand{\para}{\parallel}

\newcommand{\tr}{\mathop{\mathrm{tr}}}
\newcommand{\eqq}{~\Leftrightarrow~}

\newcommand{\bnab}{\bm{\nabla}}
\newcommand{\bk}{\bm{k}}
\newcommand{\bx}{\bm{x}}

\newcommand{\bn}{\bm{n}}

\newcommand{\bD}{\bm{D}}
\newcommand{\bB}{\bm{B}}
\newcommand{\bA}{\bm{A}}
\newcommand{\bkappa}{\bm{\kappa}}

\newcommand{\hH}{\hat{H}}
\newcommand{\hbs}{\hat{\bm{s}}}
\newcommand{\hs}{\hat{s}}

\newcommand{\hj}{\hat{j}}
\newcommand{\hi}{\hat{i}}
\newcommand{\hJ}{\hat{J}}
\newcommand{\hx}{\hat{x}}
\newcommand{\hp}{\hat{p}}

\newcommand{\rme}{\mathrm{e}}

\newcommand{\bra}[1]{\langle {#1} |}

\newcommand{\ket}[1]{| {#1} \rangle}
\newcommand{\average}[1]{\langle#1\rangle}
\newcommand{\baverage}[1]{\Big\langle#1\Big\rangle}

\newcommand{\diff}{\mathrm{d}}

\newcommand{\rmi}{\mathrm{i}}

\newcommand{\Rbb}{\mathbb{R}}

\newcommand{\with}{\quad\mathrm{with}\quad}
\newcommand{\Tr}{\mathrm{Tr}}

\newcommand{\const}{\mathrm{const.}}

\newcommand{\sgn}{\mathrm{sgn}}

\newcommand{\wl}{\mathrm{wl}}

\begin{document}

\title{
Effective field theory of magnon: \\
Dynamics in chiral magnets and Schwinger mechanism
}

\author{Masaru Hongo}
\affiliation{Department of Physics, University of Illinois, Chicago, IL 60607, USA}
\affiliation{RIKEN iTHEMS, RIKEN, Wako 351-0198, Japan}

\author{Toshiaki Fujimori}
\affiliation{Department of Physics, Keio University, Yokohama 223-8521, Japan}
\affiliation{Research and Education Center for Natural Sciences, 
Keio University, Yokohama 223-8521, Japan}

\author{Tatsuhiro Misumi}
\affiliation{Department of Mathematical Science, Akita University, Akita 010-8502, Japan}
\affiliation{Research and Education Center for Natural Sciences, 
Keio University, Yokohama 223-8521, Japan}

\author{Muneto Nitta}
\affiliation{Department of Physics, Keio University, Yokohama 223-8521, Japan}
\affiliation{Research and Education Center for Natural Sciences, 
Keio University, Yokohama 223-8521, Japan}

\author{Norisuke Sakai}
\affiliation{Department of Physics, Keio University, Yokohama 223-8521, Japan}
\affiliation{Research and Education Center for Natural Sciences, 
Keio University, Yokohama 223-8521, Japan}

\begin{abstract}
We develop the effective field theoretical descriptions 
of spin systems in the presence of symmetry-breaking effects: 
the magnetic field, single-ion anisotropy, and Dzyaloshinskii-Moriya  interaction.
Starting from the lattice description of spin systems, 
we show that the symmetry-breaking terms corresponding to the above effects can be incorporated into the effective field theory as a combination of a background (or spurious) $\SO(3)$ gauge field and 
a scalar field in the symmetric tensor representation, 
which are eventually fixed at their physical values.
We use the effective field theory to investigate mode spectra of inhomogeneous ground states, 
with focusing on one-dimensionally inhomogeneous states, such as helical and spiral states. 
Although the helical and spiral ground states share 
a common feature of supporting 
the gapless Nambu-Goldstone modes 
associated with the translational symmetry breaking, 
they have qualitatively different dispersion relations: 
isotropic in the helical phase while anisotropic in the spiral phase. 
We also discuss the magnon production induced 
by an inhomogeneous magnetic field, 
and find a formula akin to the Schwinger formula. 
Our formula for the magnon production 
gives a finite rate for antiferromagnets, 
and a vanishing rate for ferromagnets, 
whereas that for ferrimagnets interpolates 
between the two cases. 
\end{abstract}

\maketitle
\tableofcontents

\section{Introduction}

Symmetry and its spontaneous breaking give 
one of the most fundamental concepts in modern physics.
If the ground state exhibits a spontaneous breaking of continuous 
global symmetry of the system, 
the Nambu-Goldstone (NG) theorem predicts 
the inevitable appearance of the gapless excitation, 
or the NG mode~\cite{Nambu:1961tp,Goldstone:1961eq,Goldstone:1962es}.
In relativistic systems respecting the Lorentz symmetry,
the number of the broken global symmetries is matched with 
that of the NG modes 
while in nonrelativistic systems, 
it is, in general, not. 
Furthermore, in the latter case, 
dispersion relations of the resulting NG modes often 
show the quadratic behavior ($\omega = \pm a \bk^2$) 
rather than the relativistic linear one $(\omega = \pm c|\bk|)$. 
Besides, the NG modes associated with 
the spontaneous spacetime symmetry breaking 
has several characteristic behaviors such as 
the anisotropic dispersion relation realized in e.g.,
the smectic-A phase of liquid crystals~\cite{DeGennes1969,DeGennes1993,Chaikin2000} and the Fulde-Ferrell-Larkin-Ovchinnikov phase of superconductors~\cite{Fulde:1964zz,larkin:1964zz,Larkin1965}, in addition to the mismatch 
between the numbers of broken symmetries and NG modes.
Although these behaviors are beyond 
the prediction of the original NG theorem, 
the recent theoretical developments clarify 
both the counting rule and dispersion relation 
of the NG mode associated with 
the internal symmetry breaking of nonrelativistic systems~\cite{Nielsen:1975hm,Leutwyler:1993gf,Miransky:2001tw,Schafer:2001bq,Nambu:2004yia,Brauner:2010wm,Watanabe:2011ec,Hidaka:2012ym,Watanabe:2012hr,Nicolis:2012vf,Watanabe:2014fva,Takahashi:2014vua,Andersen:2014ywa,Hayata:2014yga,Beekman:2019pmi,Watanabe:2019xul}.
Also, there are several approaches to understand the NG modes for spontaneous spacetime symmetry breaking~\cite{Ivanov:1975zq,Low:2001bw,Watanabe:2013iia,Nicolis:2013sga,Hayata:2013vfa,Brauner:2014aha,Hidaka:2014fra}.
One way to work out the nonrelativistic and spacetime generalization of the NG theorem 
is to use the effective field theory (EFT)~(see e.g. Refs.~\cite{Leutwyler:1993gf,Watanabe:2012hr,Nicolis:2013sga,Watanabe:2014fva,Andersen:2014ywa}).

Magnons, or quantized spin waves, in various kinds of magnets---
antiferromagnet, ferromagnet, and ferrimagnet---give a 
canonical condensed matter example of these NG modes; relativistic one in the antiferromagnet, 
and nonrelativistic one in the ferro/ferrimagnets.
Although spin systems are originally described as lattice models, 
we can still describe their low-energy dynamics based on a continuum field theory 
at energy scales much lower than the inverse lattice spacing. 
We can regard this field theory model as an EFT of magnons that describes magnon dynamics at low energies~\cite{Burgess:1998ku,Roman1999,Hofmann1999,Bar:2003ip,Kampfer:2005ba,Gongyo:2016dzp}. 
Besides, the magnon EFT can incorporate various symmetry-breaking terms, such as a Zeeman term 
due to the coupling to external magnetic fields and single-ion anisotropy, 
as those terms are induced by small background fields 
that break symmetry explicitly (See e\,.g. Ref.~\cite{Gongyo:2016dzp}).
Thus, the spin system serves as one of the best places 
for investigating the nontrivial dynamics caused by the background field. 
An interesting symmetry-breaking term 
attracting much attention recently is the Dzyaloshinskii-Moriya 
(DM)~\cite{Dzyaloshinsky1958,Moriya:1960zz} interaction that 
arises from the spin-orbit coupling in a specific class of magnets 
called chiral magnets. 
Another example is the inhomogeneous magnetic field, 
which may drastically change the dynamics of the magnon. 

In lattice models, 
the DM interaction represents an interaction term 
proportional to the vector product of neighboring spins, 
which favors the easy-plane inhomogeneous ordering.
As a result of the competition between the DM interaction 
and the Zeeman term or the anisotropic term in the potential, 
there appears a lot of interesting inhomogeneous ground states 
such as the chiral soliton lattice in $(1+1)$-dimensional 
spin chain, and skyrmion lattice in $(2+1)$-dimensional spin 
systems, whose peculiar thermodynamic and transport behaviors 
have been recently observed~\cite{Togawa2012,Kishine2015,Togawa2016,
Muhlbauer2009,Yu2010,Heinze2011,Nagaosa2013}. 
Recent theoretical studies in the presence of the DM interaction 
have revealed various Bogomol'ny-Prasad-Sommerfield~\cite{Bogomolny:1975de,Prasad:1975kr} 
(BPS) solutions~\cite{Barton-Singer:2018dlh,Adam:2019yst,
Adam:2019hef,Schroers_2019,Ross:2020hsw} and instanton 
solutions~\cite{Hongo:2019nfr},
which enable us to study chiral magnets analytically to some extent.
When the DM interaction is more influential 
than the terms in the potential, spin systems 
tend to realize simple inhomogeneous ground states 
where the magnetization vector 
is modulated along one dimension. 
This one-dimensionally modulated ground state 
is called a helical state (see e.g., Ref.~\cite{Hongo:2019nfr}) 
or spiral state~\cite{Han_2010} 
depending on the direction of the DM interaction relative 
to the potential terms. 
In terms of the magnetization vector $\bm{n}=(n^1,n^2,n^3)$ 
with a constraint $\bm{n} \cdot \bm{n} = 1$,
we can represent the Zeeman term and the
single-ion anisotropy term as 
terms in the potential linear and quadratic 
in the third component of magnetization vector $n^3$. 
If we take the DM interaction as an interaction term
between $n^1$ and $n^2$ of spins in neighboring lattice sites, 
we find that the ground state is the helical state, 
provided the anisotropy term in the potential does not favor 
easy-axis strongly~\cite{Hongo:2019nfr}. 
If we take the DM interaction as an interaction term between $n^3$ and $n^1$, $n^2$ of spins at neighboring lattice sites, 
we find the spiral state as the ground state, 
provided potential terms are not too strong~\cite{Han_2010}. 
Thanks to its low-dimensional character, 
it is much simpler to analyze its low-energy dynamics 
than other possible inhomogeneous ground states.
Since both the helical and spiral phases show the one-dimensional
inhomogeneous order, 
we can expect that they share an essential property: 
for instance, one may expect both of them support the phonon as the NG mode associated with 
the spontaneous breaking of translational symmetry.
Nevertheless, we need to be careful 
since a general statement of the NG theorem is absent 
in the case of spontaneous breaking of the spacetime symmetry. It is worth studying the number of massless degrees of freedom and dispersion relation of these modes in each case.  

In this paper, we study the dynamics of spin systems 
by means of effective field theory, 
focusing on physical effects induced 
by explicit symmetry-breaking terms; namely 
the magnetic field, single-ion anisotropy, and the DM interaction. 
Starting from the lattice description of spin systems 
with these symmetry-breaking terms, 
we can incorporate them into the effective field theory 
by treating them as background fields 
on which the $\SO(3)$ transformation 
for the magnetization vector acts appropriately. 
We find that the DM interaction can be described 
by a background $\SO(3)$ gauge field~\cite{Schroers_2019}, 
and the magnetic field can be described 
by the temporal component of $\SO(3)$ gauge field, whereas a scalar field in the symmetric rank two tensor representation is needed to describe the single-ion anisotropy (see e.g., Ref.~\cite{Gongyo:2016dzp}). 
The assignment of the (spurious) gauge transformation rules 
to these background (spurious) fields helps 
to incorporate explicit breaking terms 
into the effective field theory of spin systems. 
This symmetry-based construction of the effective field theory 
with the DM interaction provides a unified description of magnons 
in antiferromagnets, ferromagnets, and ferrimagnets 
with the DM interaction.  

We present several applications of 
our effective field theory of magnons. 
First, we investigate the low-energy dynamics 
induced by the DM interaction. 
A simple choice of the DM interaction gives
the helical ground state, and 
another choice gives the spiral ground state. 
Both of these inhomogeneous ground states 
spontaneously break the translation symmetry 
along one direction. 
While both helical and spiral ground states 
support gapless NG modes, 
their properties are qualitatively different: 
the NG mode in the helical phase shows 
the isotropic linear dispersion relation, 
whereas that in the spiral phase shows 
the anisotropic dispersion relation. 
Moreover, the dispersion relation of the NG mode 
in the spiral phase is sensitive to the types of magnets (antiferromagnets, ferromagnets, or ferrimagnets) 
though that in the helical phase is not.
The spiral state in antiferromagnet shows 
a linear dispersion along the modulation 
and a quadratic dispersion perpendicularly. 
On the other hand, spiral states in ferromagnet and ferrimagnet show quadratic dispersions along the modulation 
and quartic dispersions perpendicularly. 
As another application, we investigate the production 
rate of magnons caused by the inhomogeneous magnetic field from the homogeneous ground state.
We show the magnon EFT with the easy-axis anisotropy can be mapped into a relativistic model of a charged scalar field, 
whose mass is determined by the sum of the easy-axis potential and the ratio of magnetization and condensation parameters.
We obtain a formula for the production rate of magnons 
analogous to the Schwinger's formula~\cite{Schwinger:1951nm}
for the charged particle pair production rate 
by a constant electric field. 
Our formula shows that the antiferromagnet corresponds to the relativistic regime (small effective mass) 
and gives the nonvanishing magnon production rate, 
whereas the ferromagnet corresponds to 
the nonrelativistic regime (infinite effective mass) 
and gives the vanishing production rate. 
The production rate for ferrimagnet interpolates 
between those for antiferromagnet and ferromagnet.

The paper is organized as follows. 
In Sec.~\ref{sec:Symmetry}, 
we describe a way to implement symmetry-breaking terms 
in EFT starting from spin systems on a lattice. 
In Sec.~\ref{sec:EFT}, we write down an EFT of magnons 
in the form of the $\O(3)$ nonlinear sigma model 
and confirm the known result for the homogeneous ground state. 
In Sec.~\ref{sec:Application}, we apply our EFT 
to the helical/spiral ground states induced by the DM interaction. 
In Sec.~\ref{sec:Pair}, we apply our EFT 
to describe the production of magnons 
by an inhomogeneous magnetic field.
Sec.~\ref{sec:Discussion} is devoted to a discussion.
In Appendix \ref{sec:Coset}, 
we present a coset construction of EFT for NG modes.

\section{Model and symmetry on lattice}
\label{sec:Symmetry}

\begin{figure}[htb]
 \centering
 \includegraphics[width=0.5\linewidth]{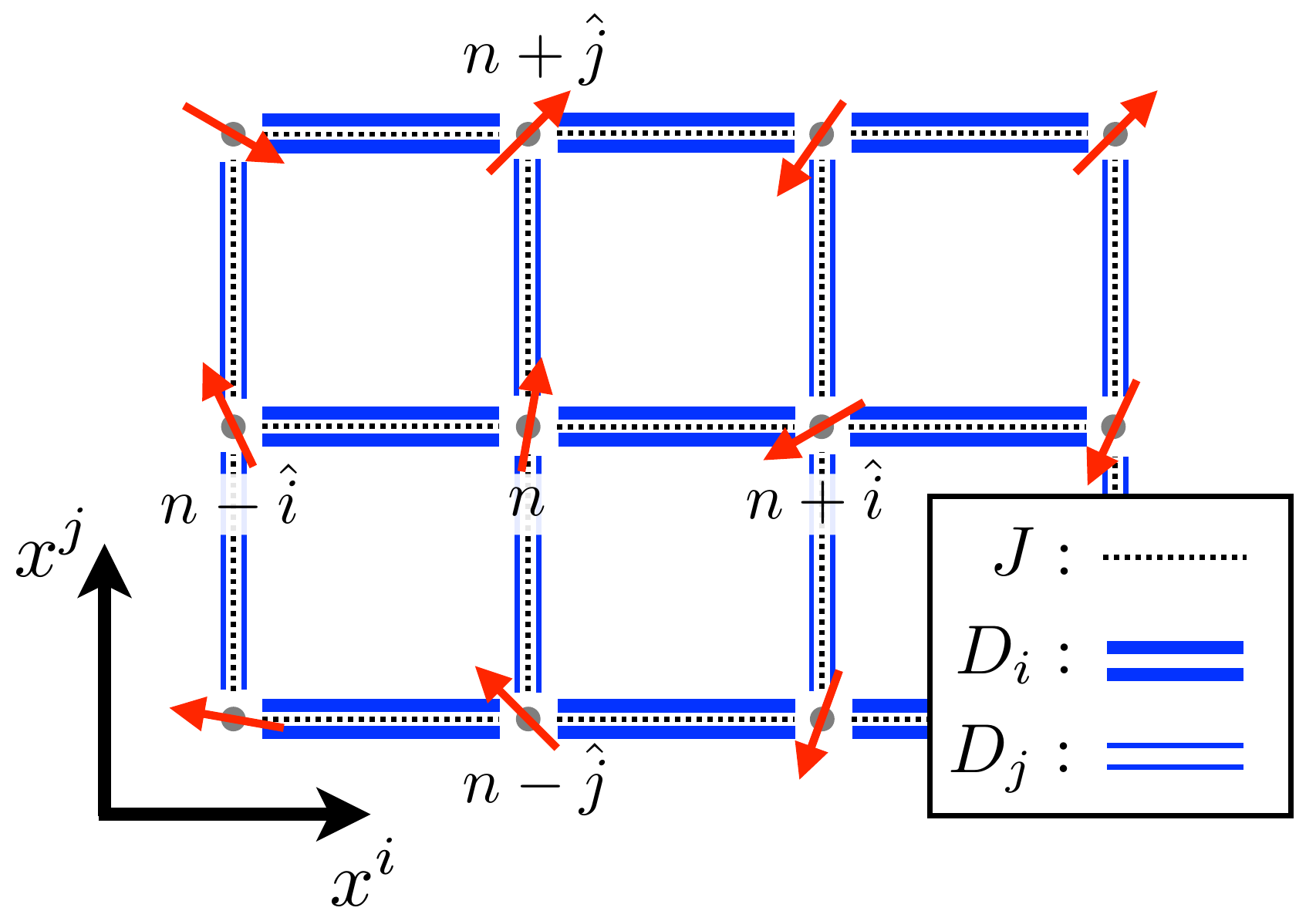}
 \caption{
 A schematic picture of the spin system under consideration. 
 We assume the localized spins live on the cubic-type lattice. 
 } 
\label{fig:Lattice}
\end{figure}

Let us consider spin systems 
whose $\SO(3)$ spin-rotation symmetry is explicitly 
but softly broken due to the external magnetic field, DM interaction, and single-ion anisotropic interaction terms.
As a concrete example, we consider Heisenberg spins, 
whose Hamiltonian reads%
\footnote{
The expression in the first line of this equation is useful when 
we explicitly consider the continuum limit. 
This is because a relation between a continuum order parameter 
field and a lattice spin vector depends on whether the system 
shows the ferromagnetic or anti-ferromagnetic order controlled 
by the sign of $J$.}
\begin{equation}
 \begin{split}
  \hH 
  &=  \sum_{n} \sum_{i=1}^d 
  \left[ 
  \frac{|J|}{2} 
  \big( \hbs^{n+\hi} - \sgn(J) \hbs^n \big)^2 + \bD_{i} \cdot 
(\hbs^n \times \hbs^{n+\hi}) 
  \right]
  - \sum_n 
  \Big[ 
  \mu \bB \cdot \hbs^n + (\hbs^n)^t C \hbs^n
  \Big]
  \\
  &= - \sum_n \sum_{i=1}^d 
  (J \delta^{ab} + D_{i}^c \epsilon_c^{~ab} ) \hs_a^n \hs_b^{n+\hi} 
  - \sum_n 
  \Big[ \mu B^a \hs^n_a + C^{ab} \hs_a^n \hs_b^{n} \Big]
  + \mathrm{(const.)}, 
 \end{split}
 \label{eq:Hamiltonian1}
\end{equation}
where $\hbs^n$ denotes a spin vector on the site $n$ 
with the (anti-)ferromagnetic interaction $J > 0\, (J<0)$, 
$\mu \bB$ the external magnetic field $\bB$ multiplied by the 
magnetic moment $\mu$, the DM interaction $\bD_i$, and anisotropic interaction $C$ known as the single-ion anisotropy.
In the second line, we have introduced the Kronecker delta $\delta^{ab}$ and the Levi-Civita symbol $\epsilon^{abc}$ 
for the internal spin indices $a,b,\cdots =1,2,3$, 
and the summation over the repeated indices is implied.
To express the nearest-neighbor pairs, 
we defined the direction $\hi = 1,2,\cdots,d$ 
with a spatial dimension $d$.
In this paper, we only consider the simple cubic-type lattice schematically shown in Fig.~\ref{fig:Lattice}, 
where the frustration in the antiferromagnetic case does not appear.
Here the DM interaction $\bD_i$ is assumed 
to have a directional dependence expressed by its subscript $i$.

In the absence of explicit symmetry-breaking terms 
($D_i^c=\mu B^a=C^{ab}$=0), 
the Hamiltonian enjoys the $\SO(3)$ symmetry, 
whose possible spontaneous breaking leads to 
the gapless collective excitations, or magnons 
(quantized spin wave) as NG modes. 
By promoting the symmetry-breaking parameters 
to background fields (spurions), 
we can construct a low-energy effective Lagrangian 
for the magnons with other possible low-energy modes 
based only on the symmetry argument~\cite{Gongyo:2016dzp}.
Although the explicit breaking terms 
break the global $\SO(3)$ symmetry, 
we can investigate their effects 
using the background field (or the so-called spurion) method
if they are small compared to the symmetric interaction 
($|D^c_i|,\, |\mu B^a|,\, |C^{ab}| \ll |J|$).

As the first attempt to parametrize the DM interaction, 
let us introduce the $\SO(3)$ gauge field coupled to 
the Noether current corresponding 
to the global $\SO(3)$ symmetry. 
When the Hamiltonian is given only by the first term in 
Eq.~\eqref{eq:Hamiltonian1}: $\hH_0 \equiv - \sum_{n,i}  J 
(\hbs^n)^t \hbs^{n+\hi} + (\mathrm{const}.)$, 
the Heisenberg equation of motion for $\hs_a^n$ 
generated by the $\SO(3)$ invariant Hamiltonian $\hH_0$ provides a 
discretized version of the conservation law of the Noether current:
\begin{equation}
 \partial_t \hJ^0_a (n)
  + \sum_{i=1}^d \big[ \hJ^i_a (n + \hi/2) - \hJ^i_a (n - \hi/2) \big] 
  = 0 
  \with
  \begin{cases}
   \hJ^0_a (n) \equiv \hs_a^n,
   \\
   \hJ^i_a (n+\hi/2) = J \epsilon_a^{~bc} \hs_b^n \hs_c^{n+\hi}.
  \end{cases}
\end{equation}
By introducing the background gauge field coupled 
to the Noether current, we obtain the following modification 
of the Hamiltonian:
\begin{equation}
 \begin{split}
  \hH_0 
  &\to \hH_0 - \sum_n A_0^a (n) \hJ^0_a (n) 
  - \sum_n \sum_{i=1}^d a A_i^a (n+\hi/2) \hJ^i_a (n+\hi/2) 
  \\
  &= - \sum_n \sum_{i=1}^d 
  \big[ J \delta^{ab} + J a A_i^c (n+\hi/2) \epsilon_c^{~ab}\big]
  \hs_a^n \hs_b^{n+\hi}
  - \sum_{n} A_0^a (n) \hs^n_a ,
 \end{split}
 \label{eq:Gauging1}
\end{equation}
with an $\SO(3)$ gauge field $A_\mu^a~(a=1,2,3)$.
We here introduced the lattice spacing $a$ 
(not the $\SO(3)$ indices) 
between spins for future convenience. 
We can now identify two symmetry-breaking terms 
$\mu B^a$ and $D_i^a$ in the original Hamiltonian 
in Eq.~\eqref{eq:Hamiltonian1} 
as the following background values of the $\SO(3)$ gauge field, 
\begin{equation}
 A_0^a (n) \equiv \mu B^a , \quad 
  A_i^a (n+ \hj/2) \equiv (Ja)^{-1} D_i^a. 
  \label{eq:BkgGauge}
\end{equation}
Although this illustrates the basic idea of the background field (spurion) method for the magnetic field and the DM interaction term, this simplified Noether procedure in Eq.~\eqref{eq:Gauging1} does not implement the local $\SO(3)$ gauge invariance fully on the lattice and is \textit{not} complete.

To correctly implement the $\SO(3)$ gauge invariance on the lattice, we draw an analogy%
\footnote{
M.H. is grateful to Takahiro Doi and Tetsuo Hatsuda for their helpful comments on lattice gauge symmetry.} 
to the Hamiltonian lattice gauge theory~\cite{Kogut:1974ag}. 
We first make the Hamiltonian $\hH_0$ invariant under the local $\SO(3)$ transformation
\begin{equation}
 \hH_0 \to 
  \hH_0' \equiv 
  \sum_{n} \sum_{i=1}^d  \frac{|J|}{2} 
  \big[ U(n,n+\hi) \hbs^{n+\hi} - \sgn (J) \hbs^n \big]^2
  - \sum_n [\bm{A}_0 (n)]^t \hbs^n ,
  \label{eq:lattice-gauge-th}
\end{equation}
where the local $\SO(3)$ transformation $g(n) \in \SO (3)$ acts 
on $\hbs^n$, $U(n,n+\hi)$, and ${\bm A}_0$ as 
\begin{equation}
 \hbs^n \to g (n) \hbs^n , \quad 
  U(n,n+\hi) \to g (n) U(n,n+\hi) g(n+\hi)^t, \quad 
  \bA_0(n) \to g(n){\bm A}_0 .
\end{equation}
Noting that the last transformation is equivalent to 
$A_0 (n) \equiv A_0^a (n) t_a \to g(n) A_0 g^t (n)$, which is identified as a time-independent gauge transformation, we will further consider time-dependent gauge invariance acting on $A_0(t,n)$ as%
\footnote{
We can rigorously justify this treatment as follows:
With the $\mathbb{C}\mathrm{P}^1$ parametrization of the spin with $z_n = (z_n^1,z_n^2)^t$, the path-integral formula gives the Lagrangian 
$  \Lcal
  = \sum_n 
  \rmi z_n^\dag \partial_0 z_n - H (s z^\dag \bm{\sigma} z)
  = \sum_n \rmi z_n^\dag D_0 z_n   - H_0 (s z^\dag \bm{\sigma} z) 
  + \sum_n \sum_{i=1}^d A_i^a J_a^i$. 
The first term with single time derivative is the so-called 
Berry phase term. We also introduced a covariant derivative 
$D_0 z_n \equiv \partial_0 z_n - \rmi sA_0^a \sigma_a z_n$ 
(corresponding to $\SO(3)$, due to the redundancy of the $\U(1)$ part $z_n \to \rme^{\rmi \theta (x)} z_n$). 
Thus, $A_0^a \hs_a^n$ term in the Hamiltonian 
comes from the correct gauging of the temporal part of $\SO(3)$ symmetry.
} 
\begin{equation}
A_0 (t,n) \to g(t,n) A_0 (t,n) g (n,t)^t 
+ \rmi g(t,n)^{-1} \partial_0 g (n,t)^t. 
\end{equation}

We now study the expansion in powers of the lattice spacing $a$ of the gauge-invariant theory \eqref{eq:lattice-gauge-th} to obtain the original Hamiltonian in 
Eq.~\eqref{eq:Hamiltonian1} with symmetry-breaking terms 
$\mu B^a,\, D_i^a$, and $C^{ab}$, ignoring higher-order terms in powers 
of $a$, which will become irrelevant in the 
continuum limit. 
By expanding the link variable $U$ at a small lattice spacing $a$, we can identify the $\SO(3)$ gauge field $A_i^a$ as 
\begin{equation}
 U(n,n+\hi) 
  = \rme^{\rmi a A_i^a (n+\hi/2) t_a}
  = I_{3\times 3} + \rmi a A_i^a (n+\hi/2) t_a + O (a^2).
\end{equation}
Counting $\hbs^{n+\hi} + \sgn (J) \hbs^n = O(a)$, the $O(a^2)$ 
term in $U$ gives higher-order terms which vanish in the continuum 
limit.
Here, we have also introduced generators of the Lie algebra $t_a \in \so(3)$ satisfying 
\begin{equation}
 [t_a, t_b] = \rmi \epsilon_{ab}^{~~c} t_c . 
\end{equation}
Using the explicit form of $t_a$ in the vector representation:
\begin{equation}
 t_1 = 
  \left(
  \begin{array}{ccc}
   0 & 0 & 0 \\[-25pt] \\
   0 & 0 & - \rmi \\[-25pt] \\
   0 & \rmi & 0
  \end{array} 
  \right), 
  \quad
 t_2 = 
  \left(
  \begin{array}{ccc}
   0 & 0 & \rmi \\[-25pt] \\
   0 & 0 & 0 \\[-25pt] \\
   - \rmi & 0 & 0
  \end{array}
  \right), 
  \quad
 t_3 = 
  \left(
  \begin{array}{ccc}
   0 & - \rmi & 0 \\[-25pt] \\
   \rmi & 0 & 0 \\[-25pt] \\
   0 & 0 & 0
  \end{array}
  \right), 
  \label{eq:so(3)-generator}
\end{equation}
we expand the gauge-invariant Hamiltonian in powers of the lattice spacing $a$ as 
\begin{equation}
 \begin{split}
  \hH_0' 
  &= \sum_{n} \sum_{i=1}^d \frac{|J|}{2} 
  \left[ 
  \hbs^{n+\hi} - \sgn (J) \hbs^n + \rmi a A_i^a t_a \hbs^{n+\hi} + O(a^2) 
  \right]^2
  - \sum_n \bA_0 \cdot \hbs_a^n
  \\
  &= \sum_{n} \sum_{i=1}^d 
  \left[
  \frac{|J|}{2} \big( \hbs^{n+\hi} - \sgn (J) \hbs^n \big)^2
  + J a \bA_i \cdot (\hbs^{n+\hi} \times \hbs^n)
  \right]
  - \sum_n 
  \left[ \bA_0 \cdot \hbs_a^n
  - \frac{|J|a^2}{2} (\hbs^{n})^t (A_i^a t_a)^2 \hbs^{n}
  \right] ,
 \end{split}
 \label{eq:Hamiltonian2}
\end{equation}
where we have not explicitly display terms 
that vanish in the naive continuum limit ($O(Ja^3)$-terms).
Comparing this with the original Hamiltonian \eqref{eq:Hamiltonian1}, 
we can confirm the identification \eqref{eq:BkgGauge} of the 
background values of the fields 
to obtain the magnetic field $\mu B$ 
and the DM interaction $D_i^a$, together with a specific value 
$C_{\mathrm{cr}}$ of the anisotropic potential [the last term $C$ in Eq.~\eqref{eq:Hamiltonian1}], given by
\begin{equation}
 C_{\mathrm{cr}} 
  = - \frac{|J|a^2}{2} (A_i^a t_a)^2 
  = - \frac{1}{2|J|} (D_i^a t_a)^2. 
  \label{eq:criticalC}
\end{equation}
This fine-tuned potential corresponds to the case of the continuum Hamiltonian whose potential can be combined with the DM interaction simply as the square of the covariant derivative.

The generic values of the single-ion anisotropy $C$ 
can also be implemented 
by introducing another background scalar field $W(n)$ 
in the symmetric rank two tensor representation, on which the local 
$\SO(3)$ transformation $g(n)$ acts as $W (n) \to g (n) W(n) g (n)^t$. 
We should identify its background value as 
\begin{equation}
 W(n) \equiv C - C_{\mathrm{cr}}  .
  \label{eq:Adjoint}
\end{equation}
Thus, apart from higher-order terms in powers of the lattice spacing 
$a$, which vanish in the na\"ive continuum limit, we find that the Hamiltonian in Eq.~\eqref{eq:Hamiltonian1} with symmetry-breaking terms $\mu B^a$, $D_i^a$ and $C^{ab}$ can be obtained from the lattice gauge invariant theory 
\begin{equation}
 \hH_0'' \equiv 
  \sum_{n} \sum_{i=1}^d  \frac{|J|}{2} 
  \big[ U(n,n+\hi) \hbs^{n+\hi} - \sgn (J) \hbs^n \big]^2
  - \sum_n A_0^a (n) \hs_a^n 
  - \sum_n  (\hbs^n)^t W(n) \hbs^n ,
  \label{eq:lattice-gauge-th2}
\end{equation}
at particular values of the background 
gauge field $A_0,\, A_i$ and scalar field $W$ given in 
Eqs.~\eqref{eq:BkgGauge} and \eqref{eq:Adjoint}.

\section{Effective Field Theory of magnons}
\label{sec:EFT}

In this section, we implement explicit symmetry-breaking terms 
presented in the previous section into a field-theoretical 
description of spin systems, or the $\O(3)$ nonlinear sigma model.
We also clarify the matching condition for the low-energy 
coefficient in the homogeneous ground state and review the low-energy 
spectrum in the absence of the explicit symmetry breaking (see also 
Appendix \ref{sec:Coset} for a coset construction as a complementary 
way to derive the effective Lagrangian).

\subsection{$\O(3)$ nonlinear sigma model description}
\label{sec:Sigma}

Since we are interested in the low-energy (long wave-length) 
behaviors of the system, we can employ the field-theoretical 
(continuum) description of the system. 
A continuum field-theoretical description of magnons (spin waves) is given 
by the $\O(3)$ nonlinear sigma model, in which a $3$-component unit vector $n^a =(n^1,n^2,n^3)^t$ with $n^a n_a = 1$ plays a role as a dynamical degree of freedom.
We note that this unit vector expresses the usual magnetization order parameter in the ferromagnetic case, while it represents the N\'eel order parameter in the antiferromagnetic case.

The local $\SO(3)$ transformation simply acts 
on the vector field $n^a$ as $\bn \to g(x) \bn$ with $g(x) \in \SO (3)$ 
as is the case with the lattice spin.
The symmetry-based discussion in the previous section enables 
us to incorporate explicit breaking terms 
in the $\O(3)$ nonlinear sigma model.
In fact, taking the continuum limit of 
the background (spurious) gauge and scalar fields 
introduced in the previous section, we 
have the $\SO(3)$ gauge field $A_\mu(x) \equiv A_\mu(x)^a t_a$ 
and the scalar $W(x)$ in the symmetric tensor representation 
on which the local $\SO(3)$ transformation 
$g(x) \in \SO(3)$ acts as 
\begin{equation}
 \begin{cases}
  A_\mu (x) \to g (x) A_\mu (x) g^{-1} (x) + \rmi g(x) \partial_\mu g^{-1} (x),
  \\
  W(x) \to g(x) W(x) g^{-1} (x) .
  \end{cases} 
  \label{eq:BkgGaugeTr}
\end{equation}
Using these, we construct 
the general local $\SO(3)$ invariant effective Lagrangian
and eventually fix the (spurious) fields to 
the nontrivial background values as 
\begin{equation}
 A_0^a (x) =  \mu B^a, \quad
  A_i^a (x) = (Ja)^{-1} D_i^a \equiv \kappa_i^a , \quad 
  W (x) = C - C_{\mathrm{cr}},
  \label{eq:Bkg-values}
\end{equation}
in order to investigate small effects of 
the explicit breaking terms in the lattice Hamiltonian \eqref{eq:Hamiltonian1}.

Using the transformation rules of the fields $n^a(x),~A_\mu^a(x)$, 
and $W(x)$ as ingredients, 
we can construct the general $\SO(3)$ 
invariant effective Lagrangian.
In the leading-order of the derivative expansion, 
where we only keep terms up to second-order in derivatives, 
the $\SO(3)$ invariant effective Lagrangian is given by 
\begin{equation}
 \Lcal_{\eff} 
  = \frac{m(n^2 \partial_0 n^1 - n^1 \partial_0 n^2)}{1+ n^3}
  + m A_0^a n_a
  + \frac{f_{\rm t}^2}{2} (D_0 n^a)^2 
  - \frac{f_{\rm s}^2}{2} (D_i n^a)^2 
  + \ell W^{ab} n_a n_b, 
  \label{eq:NLSigmaLag}
\end{equation}
where we defined a covariant derivative with the $\SO(3)$ background 
gauge field as
\begin{equation}
 D_\mu n^a \equiv \partial_\mu n^a - \epsilon^{a}_{~bc} n^b A_\mu^c.
\end{equation}
Equation~\eqref{eq:NLSigmaLag} supplemented with Eq.~\eqref{eq:Bkg-values} defines our effective field theory for general magnets including chiral magnets. 
This continuum field theory should be valid at low-energies 
and contains four parameters $m, f_{\rm t}, f_{\rm s}$, and $\ell$ as low-energy coefficients.
They can be determined from the underlying lattice model 
by the matching condition, which will be discussed shortly.
Note that the sum of the first and second term in 
Eq.~\eqref{eq:NLSigmaLag} 
manifestly breaks the Lorentz invariance%
\footnote{
More precisely, a modified Lorentz symmetry remains exact, 
see Refs.~\cite{Ohashi:2017vcy,Takahashi:2017ruq,Fujimori:2017ggv}.
} with an effective speed of light $c_s \equiv f_{\rm s}/f_{\rm t}$, 
but is $\SO(3)$ gauge invariant~\cite{Brauner:2010wm}.   
If the symmetry-breaking terms vanish ($A^a_\mu=0, W^{ab}=0$), 
the effective Lagrangian in \eqref{eq:NLSigmaLag} reduces to 
the usual $\O(3)$ nonlinear sigma model describing ferromagnets 
($m \neq 0, \,f_{\rm t}=0$), antiferromagnets $(m = 0,\,f_{\rm t}\neq 0)$, and ferrimagnets ($m \neq 0,\, f_{\rm t} \neq 0$). 
The first term is responsible for the quadratic gapless dispersion 
relation of the magnon in ferro/ferrimagnets. 
In the rest of this section, we will introduce the matching condition and study the low-energy spectrum on the top of the homogeneous ordered phase.

\subsection{Matching condition and low-energy spectrum in homogeneous order}
\label{sec:homogeneous}

Before discussing magnon dynamics in the presence of symmetry-breaking terms,
we here clarify the matching condition for low-energy coefficients $m,f_{\rm t},f_{\rm s}$ and $\ell$ in the homogeneous ground state, which breaks approximate $\SO(3)$ symmetry. 
We also demonstrate the low-energy spectrum of gapless magnons in the absence of symmetry-breaking terms.

To illustrate the procedure in a simple context, 
let us assume that the symmetry-breaking background fields gives the homogeneous ground state 
with the magnetization/N\'eel vector pointing 
the north pole as $\average{n^a} = n^a_0 \equiv (0,0,1)^t$. 
We then introduce magnon fields $\pi^\alpha~(\alpha=1,2)$ as fluctuations on the top of the ground state,
which parametrize the vector $n^a$ as 
\begin{equation}
 n^a = (\pi^1, \pi^2, \sqrt{1-(\pi^\alpha)^2})^t,
  \label{eq:parametrization-pi}
\end{equation}
where we explicitly solved the constraint $n_a n^a =1$.
Substituting this parametrization into Eq.~\eqref{eq:NLSigmaLag}, 
we obtain the effective Lagrangian of magnons given by 
\begin{align}
 \Lcal_{\eff}
 &= - \frac{m}{2} \epsilon^3_{~\alpha\beta} \pi^\alpha \partial_0 \pi^\beta
 + m \left( 
 \delta^3_a + \epsilon^3_{~\alpha a} \pi^\alpha 
 - \frac{1}{2} \delta^3_a (\pi^\alpha)^2
 \right) A_0^a
 + \ell W^{33}
 \nonumber
 \\
 &\quad + \frac{f_{\rm t}^2}{2} \delta_{\alpha\beta} 
D_0 \pi^\alpha D_0 \pi^\beta
 - \frac{f_{\rm s}^2}{2}\delta^{ij} \delta_{\alpha\beta} 
D_i \pi^\alpha D_j \pi^\beta
 - \ell \left[ (W^{23} + W^{32}) \pi^1 - (W^{13} + W^{31}) \pi^2 \right]
 \nonumber
  \\
  &\quad
  - \ell
  \left[ (W^{33} - W^{22}) (\pi^1)^2
 + (W^{33} - W^{11}) (\pi^2)^2
 + (W^{12} + W^{21}) \pi^1 \pi^2
  \right]
  + \Lcal_{\mathrm{int}},
 \label{eq:EffLag1}
\end{align}
where $\Lcal_{\mathrm{int}}$ contains more than two magnon fields 
representing interactions between them.
We have also defined the covariant derivative of the magnon field as
\begin{equation}
 D_\mu \pi^\alpha \equiv 
  \partial_\mu \pi^\alpha - A_\mu^\alpha 
  - A_\mu^3 \epsilon^\alpha_{~\beta3} \pi^\beta .
\end{equation}
Since the ground state spontaneously breaks the $\SO(3)$ symmetry down to 
its subgroup $\SO(2)_z$,  the magnon fields can be identified as the NG bosons. 
The effective Lagrangian \eqref{eq:EffLag1} is reparametrized by $\pi^\alpha$ fields in order to make their 
role as NG bosons manifest, and is equivalent to the original 
effective field theory in Eq.~\eqref{eq:NLSigmaLag}, provided 
all order terms of $\pi^\alpha$ in $\Lcal_{\mathrm{int}}$ are kept. 
In the rest of this section, we assume symmetry-breaking terms in Eq.~\eqref{eq:EffLag1} do not induce a tachyon-like instability around the assumed ground state $n^a_0 \equiv (0,0,1)^t$.
Thus, the actual values of the symmetry-breaking terms in Eq.~\eqref{eq:EffLag1} cannot be arbitrary.

We now discuss the matching conditions within a tree-level 
analysis in order to fix the phenomenological parameters in the 
effective Lagrangian \eqref{eq:EffLag1}: the four parameters  
$m,~f_{\rm t},~f_{\rm s}$ and $\ell$. 
We first introduce the $\SO(3)$ current 
defined by the variation of the effective action $\Scal$ 
in terms of the $\SO(3)$ gauge fields 
\begin{equation}
 J^\mu_a (x) = 
  \frac{\delta \Scal [\pi^\alpha;A_\mu^a,W]}{\delta A_\mu^a (x)} 
  \with
  \Scal [\pi^\alpha;A_\mu^a,W] 
  = \int \diff^{d+1} x \, \Lcal_{\eff}.
\end{equation}
Up to quadratic terms in the magnon field $\pi^\alpha(x)$, 
the $SO(3)$ currents are explicitly given by
\begin{equation}
\begin{split}
 J^0_3
 &= m - \frac{m}{2} (\pi^\alpha)^2  
 - f_{\rm t}^2 \epsilon_{\alpha\beta3} \pi^\beta
 ( \partial_0 \hat{\pi}^\alpha - A_0^\alpha 
 - A^3_0 \epsilon^\alpha_{~\gamma3} \pi^\gamma )
 + \cdots ,
 \\
 J^i_3 
 &= f_{\rm s}^2 \delta^{ij} \epsilon_{\alpha\beta3} \pi^\beta
 ( \partial_j \pi^\alpha - A_j^\alpha 
 - A^3_j \epsilon^\alpha_{~\gamma3} \pi^\gamma )
 + \cdots , 
 \\
 J^0_\alpha
 &= - m \epsilon^3_{~\alpha\beta} \pi^\beta
 - f_{\rm t}^2 \delta_{\alpha\beta}
 ( \partial_0 \pi^\beta - A_0^\beta 
 - A^3_0 \epsilon^\beta_{~\gamma3} \pi^\gamma )
 + \cdots ,
 \\
 J^i_\alpha
 &= f_{\rm s}^2 \delta^{ij} \delta_{\alpha\beta}
 ( \partial_j \pi^\beta - A_j^\beta 
 - A^3_j \epsilon^\beta_{~\gamma3} \pi^\gamma )
 + \cdots.
\end{split}
\label{eq:GaugeCurrent}
\end{equation}
We then define the generating functional $Z[A_\mu^a,W]$
for the $\SO(3)$ current $J_a^\mu$ by the path integral over $\pi^\alpha(x)$
\begin{equation}
 Z[A_\mu^a,W] = \int \mathcal D \pi^\alpha \,
  \exp \left( \rmi \Scal [\pi^\alpha;A_\mu^a,W] \right). 
  \label{eq:generating_functional}
\end{equation}
The expectation values of the currents can be obtained 
by taking the functional derivative with respect to $A_\mu^a$ 
\begin{equation}
 \average{J_a^\mu (x)} 
  = \rmi^{-1} \frac{\delta}{\delta A_\mu^a (x)} \log Z [A_\mu^a,W].
\end{equation}
We can also introduce the generalized susceptibility 
for the $\SO(3)$ symmetry (correlation functions 
of current operators), defined by
\begin{equation}
 \chi^{\mu\nu}_{ab} (\omega,\bk) 
  = - \int \diff^{d+1} x \, \rme^{\rmi \omega t - \rmi \bk \cdot \bx } 
  \frac{\delta}{\delta A_\mu^a(x)} \frac{\delta}{\delta A_\nu^b (0)} \log Z[A_\mu^a,W].
  \label{eq:Suscept}
\end{equation} 
If we wish to find the expectation values of 
the current operators 
and the susceptibility at the tree level, 
we just need to evaluate Eq.\,\eqref{eq:generating_functional} 
at the homogeneous ground state $\pi^\alpha(x)=0$ and 
the background values of $A_\mu^a$ and $W(x)$. 
Denoting the ground state expectation value of 
an operator $O$ by $\average{O}$, 
we obtain the expectation value of 
the current operator $\hat J_a^\mu$ 
and the correlation functions of the current operators 
(susceptibility) at the tree level approximation as 
\begin{equation}
 \begin{cases}
  \left. \average{J_3^0 (x)} \right|_{\pi = 0} 
  = m, 
  \\   
  \left. \chi_{\alpha\beta}^{00} (\omega=0,\bk=0) \right|_{\pi = 0} 
  = f_{\rm t}^2 \delta_{\alpha\beta}, 
  \\
  \left. \chi_{\alpha\beta}^{ij} (\omega=0,\bk=0) \right|_{\pi = 0} 
  = - f_{\rm s}^2 \delta^{ij} \delta_{\alpha\beta}.
 \end{cases}
 \label{eq:matcthing}
\end{equation}
Throughout this section, 
we use an abbreviated notation of 
$\pi^\alpha=0$ to denote the ground state values: 
$\pi^\alpha=0$ and the background field
$A_\mu^a$ and $W$ fixed at physical values given in Eq.~\eqref{eq:Bkg-values}. 
The second and third equations indicate 
a nonvanishing long-range correlation 
for the (approximately) conserved currents, 
which is a manifestation of 
spontaneous symmetry breaking. 
Its structure is the same as the familiar symmetry breaking 
in the Lorentz invariant systems,
except for the independent numerical prefactor, 
which reflects the fact that 
the propagating speed of magnons is 
generally not the speed of light.
On the other hand, the first equation is peculiar to 
the nonrelativistic system 
since the nonvanishing charge density $m \neq 0$ 
manifestly breaks the Lorentz invariance.

Taking the variation with respect to the background field $W$, 
we can also obtain the matching condition for $\ell$ as
\begin{equation}
 \ell = 
  \left. \baverage{ \frac{\delta \Scal_{\eff}}{\delta W^{33}} }
  \right|_{\pi=0},
  \label{eq:matcthing2}
\end{equation}
which is proportional to $\average{\hs_3^n \hs_3^n}$ 
in the lattice model description.
Equations~\eqref{eq:matcthing}-\eqref{eq:matcthing2} provide the matching condition for low-energy coefficients $m,f_{\rm t},f_{\rm s}$ and $\ell$.

Depending on which coefficients are present, we can classify 
various magnets into three types: antiferromagnets, ferromagnets, 
and ferrimagnets.
For simplicity, let us consider the simple situation with vanishing 
explicit symmetry-breaking terms---the background magnetic field $B^a$, 
DM interaction $D_j^a$, and single-ion anisotropy $C^{ab}$. 
In this case, we can simplify the quadratic part of the effective 
Lagrangian as
\begin{align}
 \Lcal_{\eff}^{(2)}
 &= - \frac{m}{2} \epsilon^3_{~\alpha\beta} \pi^\alpha \partial_0 \pi^\beta
 + \frac{f_{\rm t}^2}{2} \delta_{\alpha\beta} 
\partial_0 \pi^\alpha \partial_0 \pi^\beta
 - \frac{f_{\rm s}^2}{2} \delta_{\alpha\beta} \delta^{ij} 
 \partial_i \pi^\alpha \partial_j \pi^\beta,
 \label{eq:EffLag-quadratic}
\end{align}
which results in the following equation of motion:
\begin{equation}
 \begin{pmatrix}
  f_{\rm t}^2 \partial_0^2 - f_{\rm s}^2 \bnab^2 & m \partial_0  \\
  - m \partial_0 & f_{\rm t}^2 \partial_0^2 - f_{\rm s}^2 \bnab^2
 \end{pmatrix}
 \begin{pmatrix}
  \pi^1 \\
  \pi^2
 \end{pmatrix}
 = 0 .
 \label{eq:eom-homogeneous}
\end{equation}
By solving the characteristic equation for the coefficient matrix, 
we can investigate the number of the independent NG modes and 
their dispersion relations.
The result is summarized as follows:
\begin{equation}
 \begin{split}
  &\bullet~\mbox{Antiferromagnet}~
  (f_{\rm t}\neq 0, ~m = 0 ) :~
  2\, \mbox{NG modes} \with 
  \omega = c_s |\bk| ,
  \\
  &\bullet~\mbox{Ferromagnet}~
  (f_{\rm t} = 0,~m \neq 0) \hspace{20pt} :~
  1\, \mbox{NG mode} \ \with 
  \omega = \frac{f_{\rm s}^2}{m}\bk^2 ,
  \\
  &\bullet~\mbox{Ferrimagnet}~
  (f_{\rm t} \neq 0,~m \neq 0) \hspace{22pt} :~
  1\, \mbox{NG mode and } 1\, \mbox{gapped mode} 
  \\
  &\hspace{180pt}\mathrm{with}\quad
  \omega = \dfrac{f_{\rm s}^2}{m}\bk^2 + O(\bk^4) , \quad
   \omega = \dfrac{m}{f_{\rm t}^2} + O (\bk^2). 
 \end{split}
 \label{eq:DR1}
\end{equation}
We list only positive frequencies here and subsequently, although 
there are negative frequency solutions with the opposite sign. 
Here, we have introduced the propagating speed of the antiferromagnetic 
magnon as $c_s \equiv f_{\rm s}/f_{\rm t}$, 
which is not necessarily 
the speed of light in contrast to the NG mode 
in the Lorentz invariant system.
Note that the ferro/ferrimagnetic magnons show the quadratic 
dispersion relation, and the number of gapless excitation 
$N_{\mathrm{NG}}$ obeys the general counting rule 
\begin{equation}
 N_{\mathrm{NG}} = N_{\mathrm{BS}} - \mathrm{rank} \, \rho 
  \with
  \rho_{ab} (x) \equiv \average{[\rmi \hat{Q}_a, \hat{J}^0_b (x)]}
\end{equation}
Here, we introduced the number of the broken symmetry $N_{\mathrm{BS}}$ and the so-called Watanabe-Brauner matrix $\rho_{ab}$, where $\hat{Q}_a = \int \diff^d x \, \hat{J}^0_{a} (x)~(a=1,2,3)$ denotes the Noether charge associated with the $\SO(3)$ symmetry.
This result is completely consistent with 
the above matching condition 
since $\rho_{12} (x) = \average{\hJ^0_3 (x)} = m$ does not vanish in the homogeneous ferro/ferrimagnetic ground state while it does in the antiferromagnetic one.
The dispersion relation at small $\bk$ in the ferrimagnet case given in Eq.~\eqref{eq:DR1} reduces to that in the ferromagnet case as $f_{\rm t}\to 0$
while it does not reduce to that of antiferromagnet in the limit of $m\to 0$.
This apparent inconsistency comes from our limiting procedure:
we first take small $\bk$ limit in Eq.~\eqref{eq:DR1} and consider $m \to 0$ limit.
The full dispersion relation for ferrimagnets is available from Eq.~\eqref{eq:eom-homogeneous}, which, of course, reproduces that of antiferromagnets when we take $m \to 0$.

\section{Low-energy spectrum on helical/spiral phase}
\label{sec:Application}

In this section, we apply the effective Lagrangian~\eqref{eq:NLSigmaLag} 
to study low-energy excitation spectra of 
inhomogeneous ground states induced by the DM interaction. 
When the DM interaction is sufficiently large, 
inhomogeneous states tend to become the ground state. 
The simplest of such inhomogeneous ground states develop 
a one-dimensional modulation of the spin vector. 
Depending on the types of the DM interaction, 
they are called the helical ground state or spiral ground state. 
Both of them support a gapless NG mode 
as a low-energy excitation, 
that is, the phonon associated with 
the spontaneous breaking of the translation symmetry.
Nonetheless, we demonstrate that 
the form of the dispersion relation 
is qualitatively different 
between helical and spiral states.

\subsection{Isotropic dispersion relation in helical ground state}
\label{sec:helical}

As the first application, we consider the case 
where the simple combination of the uniaxial DM interaction 
and easy-axis anisotropic potential 
along the same direction are present. 
For simplicity, we choose the following background values 
for the external fields in the effective Lagrangian~\eqref{eq:NLSigmaLag}:
\begin{equation}
 A_0^a = 0, \quad A_i^a = \kappa_i \delta^a_3, \quad 
  \mathrm{and} \quad 
 \ell W^{ab} = \frac{W}{2} \delta^a_3 \delta^b_3 .
  \label{eq:DM-helical}
\end{equation}
Using this setup, we will show that the system develops the helical 
order in the case of the easy-plane potential ($W<0$)~\cite{Hongo:2019nfr}.
Due to the spontaneous breaking of the translational symmetry, 
the helical ground state is shown to support a translational gapless phonon (NG mode) 
in the low-energy spectrum irrespective 
of the types of chiral magnets.

In contrast to the analysis in Sec.~\ref{sec:homogeneous}, 
we have an inhomogeneous ground state, 
which forces us to abandon the description in Eq.~\eqref{eq:EffLag1} 
in terms of NG bosons on the homogeneous ground state.
We thus here start with the original $\O(3)$ nonlinear sigma model description given in Eq.~\eqref{eq:NLSigmaLag}.
Substituting Eq.~\eqref{eq:DM-helical} 
into the effective Lagrangian 
\eqref{eq:NLSigmaLag}, we obtain 
\begin{equation}
 \Lcal_{\eff} 
  = \frac{m(n^2 \partial_0 n^1 - n^1 \partial_0 n^2)}{1+ n^3}
  + \frac{f_{\rm t}^2}{2} (\partial_0 n^a)^2 
  - \frac{f_{\rm s}^2}{2} (\partial_i n^a - \kappa_i \epsilon^a_{~b3} n^b )^2 
  + \frac{W}{2} (n^3)^2.
  \label{eq:NLSigmaLag2}
\end{equation}
One should note that the potential $V(n^3)=-W(n^3)^2/2$ favors 
$n^3=\pm 1$ (easy-axis) if $W>0$, whereas it favors $n^3=0$ 
(easy-plane) if $W<0$. 
To find the ground state, we use the Hamiltonian defined by the 
Legendre transformation of Eq.~\eqref{eq:NLSigmaLag2} as 
\begin{equation}
 \begin{split}
  \Hcal
  &= \Pi_a \partial_0 n^a - \Lcal_{\eff} 
  \\
  &= \frac{1}{2f_{\rm t}^2}
  \left( \Pi_a - \frac{m (n^2 \delta_a^1 - n^1 \delta_a^2)}{1+ n^3} \right)^2
  + \frac{f_{\rm s}^2}{2} (\partial_i n^a - \kappa_i \epsilon^a_{~b3} n^b )^2 
  - \frac{W}{2} (n^3)^2.
  \\
 \end{split}
 \label{eq:EffH}
\end{equation}
where we defined the conjugate momentum $\Pi_a$ as 
\begin{equation}
 \Pi_a \equiv \frac{\partial \Lcal_{\eff}}{\partial (\partial_0 n^a)}
  = \frac{m (n^2 \delta^1_a - n^1 \delta^2_a )}{1+n^3} 
  + f_{\rm t}^2 \partial_0 n_a.
\end{equation}
Noting that the Hamiltonian \eqref{eq:EffH} is expressed 
as a sum of the quadratic terms, 
we try to find a candidate ground state solution 
by requiring the first two terms to vanish:
\begin{equation}
 \partial_0 n^a =0 , \quad 
  \partial_i n^a - \kappa_i \epsilon^a_{~b3} n^b = 0.
\end{equation}
The solutions of this set of equations are given by
\begin{equation}
 \bar{n}^a = 
  \begin{pmatrix}
   \sqrt{1- \bar{A}^2} \cos (\bkappa \cdot \bx + \bar{\phi}) \\
   - \sqrt{1 - \bar{A}^2} \sin (\bkappa \cdot \bx + \bar{\phi}) \\
   \bar{A}
  \end{pmatrix},
  \label{eq:helical-config}
\end{equation}
where two real parameters $\bar{A} \in [-1,1]$ and $\bar{\phi} \in [0,2\pi)$ 
denote integration constants, which characterize an orbit 
on the unit sphere at a constant latitude $n^3=\bar A$. 
Since we can regard the potential to be a function of these 
orbits ($n_3=\bar A$), we can find the ground state by 
just finding the orbit corresponding to the minimum of the potential. 
Thus, we find the ground states as 
\begin{equation}
 \bar{A} = \left\{
  \begin{array}{c l}
   \pm 1 \quad & \mathrm{for} \quad W > 0 ,
   \\
 0 \quad  & \mathrm{for} \quad W < 0,
   \\
   \mathrm{arbitrary} \in [-1,1] \quad & \mathrm{for} \quad W=0.
  \end{array}
  \right.
  \label{eq:A-cond}
\end{equation}

\begin{figure}[b]
 \centering
 \includegraphics[width=0.5\linewidth]{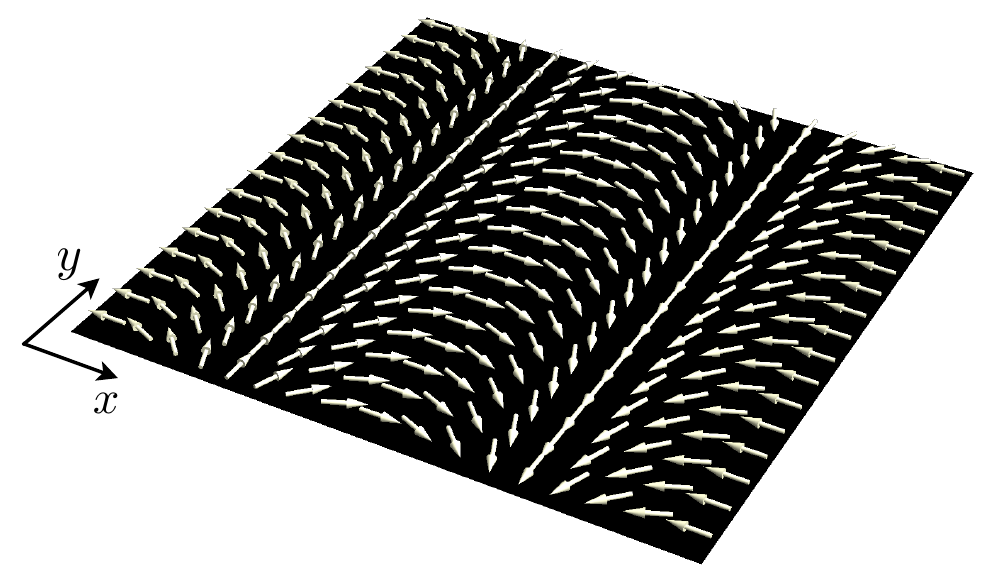}
 \caption{
A schematic picture of the helical ground state 
in the DM dominant $(2+1)$-dimensional magnet.
 } 
\label{fig:helical-config}
\end{figure}

While the ground state is homogeneous  for $\bar{A} = \pm 1$, 
it realizes the inhomogeneous helical order for $|\bar{A}| < 1$.
Figure \ref{fig:helical-config} shows a schematic picture of 
the helical ground state configuration of $n^a$ with $\bar{A} = 0$ 
(the orbit circling at the equator).
Thus, we find that the helical order is realized along 
the direction of the DM interaction $\bkappa$ 
when $W \le 0$ (see also Ref.~\cite{Hongo:2019nfr}).
The fine-tuned case with $W=0$ is unique in the sense that
circles at any latitude give the degenerate classical 
ground states corresponding to 
the Kaplan-Shekhtman-Aharony-Entin-Wohlman
limit~\cite{KSEA1,KSEA2}.
In this case of $W=0$, BPS soliton solutions in ($1+1$)-dimension have been 
exhaustively worked out in Ref.~\cite{Hongo:2019nfr}. 
One should note that our parametrization of the $(n^3)^2$ term in the potential differs from many previous works, including Ref.~\cite{Hongo:2019nfr}, 
where the additional term was present in the potential as 
$V(n^3)=(-W+\bkappa^2)(n^3)^2/2$.

Let us then consider the case with $W<0$, 
and investigate the low-energy spectrum 
on the helical ground state.
For that purpose, we consider fluctuations of 
$\delta A$ and $\delta \phi$ 
around the fixed background values 
$\bar{A} = 0$ and $\bar{\phi} = 0$, 
and rewrite the effective Lagrangian 
by promoting $\delta A(x)$ and $\delta \phi(x)$ 
to dynamical fields.
In short, we parametrize the spin vector $n^a$ as
\begin{equation}
 n^a(x) = 
  \begin{pmatrix}
   \sqrt{ 1 - (\delta A(x) )^2 } \cos (\bkappa \cdot \bx + \delta \phi(x))
   \\
   - \sqrt{1- (\delta A(x) )^2} \sin (\bkappa \cdot \bx + \delta \phi(x))
   \\
   \delta A(x)
  \end{pmatrix}
  \simeq 
  \begin{pmatrix}
   \phantom{-} \cos (\bkappa \cdot \bx + \delta \phi(x))
   \\
   - \sin (\bkappa \cdot \bx + \delta \phi(x))
   \\
   \delta A(x)
  \end{pmatrix},
  \label{eq:helical-config}
\end{equation}
where, in the second equality, we have retained the leading 
order of the expansion with respect to the fluctuation ($\delta A$ 
and $\delta \phi$) to investigate the low-energy spectra of  
$\delta A$ and $\delta \phi$.
Substituting this parametrization into the original effective 
Lagrangian \eqref{eq:NLSigmaLag2}, 
we now obtain the quadratic 
part of the Lagrangian 
for the amplitude mode $\delta A$ and 
phase mode $\delta \phi$ as 
\begin{equation}
 \begin{split}
  \Lcal_{\eff}^{(2)}
  &= m (1 - \delta A) \partial_0 \delta \phi
  + \frac{f_{\rm t}^2}{2} 
  [ (\partial_0 \delta A)^2 + (\partial_0 \delta \phi)^2 ]
  - \frac{f_{\rm s}^2}{2} 
  [(\partial_i \delta A)^2 + (\partial_i \delta \phi)^2 ]
  + \frac{W}{2} (\delta A)^2,
 \end{split}
\label{eq:quadratic_lag}
\end{equation}
from which we can read off the following linearized equations of motion: 
\begin{equation}
 \begin{pmatrix}
  f_{\rm t}^2 \partial_0^2 - f_{\rm s}^2\bnab^2 & - m \partial_0 \\
  m \partial_0 & f_{\rm t}^2 \partial_0^2 - f_{\rm s}^2\bnab^2 -W
 \end{pmatrix}
 \begin{pmatrix}
  \delta \phi 
  \\
  \delta A
 \end{pmatrix}
 = 0 .
 \label{eq:eom-helical}
\end{equation}
Note that the equations of motion for the amplitude and phase 
fluctuations are coupled in the presence of the magnetization 
parameter $m$ while they decouple for vanishing $m$.
Solving the characteristic equation for the matrix 
and noting $W=-|W|$, 
we obtain the dispersion relation in each case of magnets as 
\\ 
$\bullet~\mbox{Antiferromagnet}~
  (f_{\rm t} \neq 0,~m = 0):$
  \begin{align}
  \omega ~=~ \frac{f_{\rm s}}{f_{\rm t}} |\bk|  , \quad 
 \frac{\sqrt{|W| + (f_{\rm s}\bk)^2}}{f_{\rm t}},
   \label{eq:DR-anti-W} 
  \end{align}
  $\bullet~\mbox{Ferromagnet}~
  (f_{\rm t} = 0,~m \neq 0):$
  \begin{align}
   \omega ~=~ 
   \frac{f_{\rm s}|\bk| \sqrt{|W| + (f_{\rm s}\bk)^2}}{m},
  \label{eq:DR3} 
  \end{align}
$\bullet~\mbox{Ferrimagnet}~
  (f_{\rm t} \neq 0,~m \neq 0):~$
\begin{align}
  \omega ~=~ 
  \begin{cases}
   \left(\dfrac{|W|}{m^2 + |W|}\right)^{\frac{1}{2}} 
   \dfrac{f_{\rm s}}{f_{\rm t}} |\bk| 
   +\dfrac{m^4 }{2\sqrt{|W|(m^2+|W|)^5}}
   \dfrac{\left(f_{\rm s} |\bk|\right)^3}{f_{\rm t}} 
   + O(|\bk|^5) , 
   \vspace{5pt} \\
   \dfrac{\sqrt{m^2 +  |W|}}{f_{\rm t}} 
   + \dfrac{2m^2+|W|}{2(m^2 + |W|)^{3/2}} \dfrac{f_{\rm s}^2 \bk^2}{f_{\rm t}}
+ O(\bk^4).
   \\
  \end{cases}
 \label{eq:DR-ferri-W}
\end{align}
We now see that the system supports only one gapless excitation 
(NG mode), and its dispersion relation is 
linear with respect to the momentum in all cases, 
in contrast to the case of $W=0$ in Eq.~\eqref{eq:DR-ferri-W}. 

In the current case ($W<0$), the dispersion relation 
at small $\bk$ in the ferrimagnet case \eqref{eq:DR-ferri-W}
reduces to that in the antiferromagnet case \eqref{eq:DR-anti-W}
in the limit of $m\to 0$, 
but does not reduce to that in the ferromagnet case of \eqref{eq:DR3} 
in the limit of $f_{\rm t}\to 0$. 
On the other hand, the dispersion relations for antiferromagnet in 
Eq.~\eqref{eq:DR-anti-W} and ferromagnet in Eq.~\eqref{eq:DR3} 
reduce in the limit $W \to 0$ to Eq.~\eqref{eq:DR1} 
for anti-ferromagnet and ferromagnet. 
However, the dispersion relation of the gapless mode of ferrimagnet in Eq.~\eqref{eq:DR-ferri-W} is 
singular as $W\to 0$ and does not agree with Eq.~\eqref{eq:DR1}. 
This discontinuity is due to the change of the small $\bk$ behavior 
from $|W||\bk|$ at $W<0$ to $\bk^2$ at $W=0$, which is similar to magnon dispersion relations in the homogeneous ordered phase. 
Again, the full dispersion relation for ferrimagnets before small-$\bk$ expansion, which is available from solving \eqref{eq:eom-helical}, reproduces both ferromagnetic and antiferromagnetic limits.

One may regard this gapless mode as the translational phonon. 
However, we note that it is also possible to interpret this mode as the magnon mode in a ``rotating frame''.
This is because one can eliminate the DM interaction by performing the field redefinition of the spin vector (see e.g., Ref.~\cite{Hongo:2019nfr}). 
As a result, the newly defined spin develops the homogeneous order so that one obtains the corresponding magnon mode.
In this interpretation, the linear dispersion with 
$W < 0$ corresponds to the magnon 
in the presence of the easy-plane potential, 
where the remaining $\SO(2)$ symmetry is spontaneously 
broken (the spectrum with $W = 0$ corresponds to the magnon 
without $\SO(3)$ symmetry breaking perturbations).

\subsection{Anisotropic dispersion relation in spiral ground state}
\label{sec:Spiral}

Let us next consider the $(2+1)$-dimensional chiral magnets 
containing an isotropic (in the $x, y$ plane) DM interaction 
$D_i^a \propto \delta_i^a, ~ i=1,2$, 
which allows a spiral ground state 
when the DM interaction is more dominant
than the potential. 
In order to obtain the simplest explicit solution, 
we consider the case without a potential 
(after the DM interactions are explicitly 
separated from covariant derivatives). 
Hence our effective Lagrangian is given by 
\begin{equation}
 \Lcal_{\eff} = \frac{m (n^2 \partial_0 n^1 - n^1 \partial_0 n^2)}{1+n^3} 
  + \frac{f_{\rm t}^2}{2} (\partial_0 n^a)^2 
- \frac{f_{\rm s}^2}{2} (\partial_i n^a)^2
  + f_{\rm s}^2\kappa 
  \big[
  n^3 (\partial_y n^1 - \partial_x n^2) + (n^2 \partial_x - n^1 \partial_y) n^3
  \big] .
\label{eq:spiral-lag1}
\end{equation}
This Lagrangian corresponds to the following choice of background fields 
with a constant term $-\kappa^2/2$ discarded: 
\begin{equation}
 A_0^a = 0, \quad 
  A_i^a = - \kappa \delta_i^a, \quad
  \ell W_{ab} n^a n^b = - f_{\rm s}^2\kappa^2 (n^3)^2.
\end{equation}
Instead of treating the constrained variable $n^a$, we now 
explicitly solve the constraint $n^a n_a = 1$ by using the 
spherical parameterization of the spin vector $n^a$ given by
\begin{equation}
n^a = (\sin \theta \cos \phi, \sin \theta \sin \phi, \cos \theta)^t
 \with 0 \leq \theta < \pi, \quad 0 \leq \phi < 2\pi.
\end{equation}
Substituting this into Eq.~\eqref{eq:spiral-lag1}, 
we obtain the Lagrangian in terms of the unconstrained variables
\begin{equation}
 \begin{split}
  \Lcal_{\eff} 
  =& 2 m \sin^2 \frac{\theta}{2} \partial_0 \phi 
  + \frac{f_{\rm t}^2}{2} 
  \big[ 
  (\partial_0 \theta)^2 + \sin^2 \theta (\partial_0 \phi)^2
  \big]
  - \frac{f_{\rm s}^2}{2} 
  \big[ 
  (\partial_i \theta)^2 + \sin^2 \theta (\partial_i \phi)^2
  \big]
  \\
  &+ f_{\rm s}^2\kappa 
  \Big[ 
  ( \cos \phi \partial_y -  \sin \phi \partial_x) \theta 
  - \frac{1}{2} \sin 2 \theta 
  (\cos \phi \partial_x + \sin \phi \partial_y) \phi
  \Big] .
 \end{split}
 \label{eq:Lagrangian-isotropic}
\end{equation}

\begin{figure}[b]
 \centering
 \includegraphics[width=0.5\linewidth]{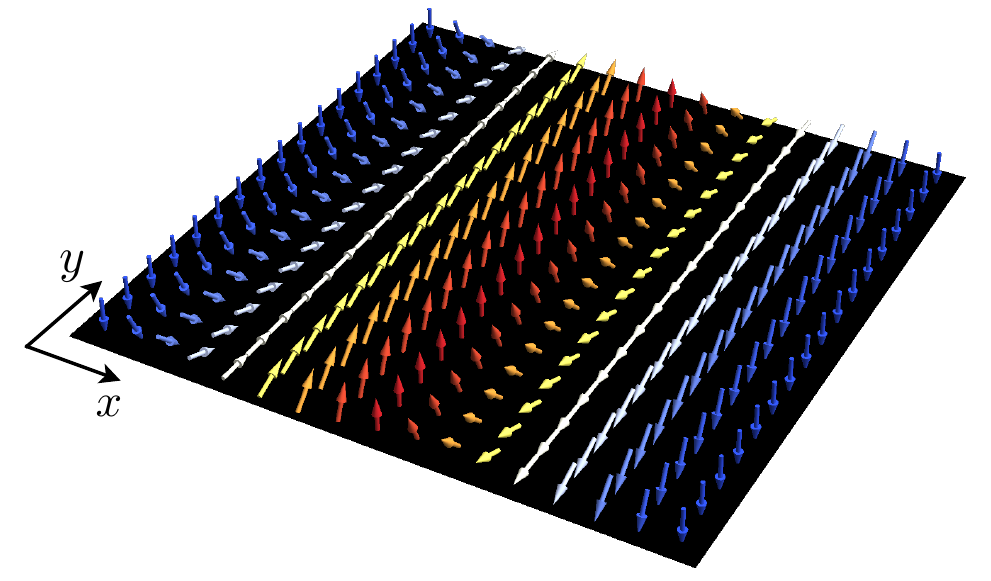}
 \caption{
 A schematic picture of the spiral ground state in the 
 $(2+1)$-dimensional magnet.
 } 
\label{fig:spiral-config}
\end{figure}

To find a one-dimensionally inhomogeneous solution, 
let us assume that the configuration is independent of time $t$ and spatial coordinate $y$. 
This assumption is consistent with the equation of motion, 
thanks to the spacetime translational symmetry. 
Retaining only the $x$-dependence, 
we find energy density ${\cal E}$ of such a configuration as
\begin{equation}
 {\cal E} [\theta, \phi] =
 \frac{f_{\rm s}^2}{2} 
 \left[ 
  \left(\frac{\diff \theta}{\diff x}\right)^2 
  + \sin^2 \theta \left(\frac{\diff \phi}{\diff x}\right)^2
 \right]
 - f_{\rm s}^2\kappa 
 \left[ - \sin \phi \frac{\diff \theta}{dx} 
  - \frac{1}{2} \sin 2 \theta 
  \cos \phi \frac{\diff \phi}{\diff x} \right] .
 \label{eq:energyDensity1}
\end{equation}
The equation of motion for $\phi$ can be solved trivially by taking 
\begin{equation}
 \phi=\pm\frac{\pi}{2} + 2n \pi, \quad n\in Z . 
\end{equation}
With this choice, the energy density becomes 
\begin{equation}
 {\cal E}[\theta]=
  \frac{f_{\rm s}^2}{2} 
  \left(\frac{\diff \theta}{\diff x}\right)^2 
  \pm f_{\rm s}^2\kappa \frac{\diff \theta}{\diff x} , 
 \label{eq:energyDensity2}
\end{equation}
where $\pm$ sign corresponds to $\phi=\pm \pi/2+2n\pi$. 
It is interesting to observe that the DM interaction for the 
one-dimensionally inhomogeneous configuration becomes a total 
derivative and does not affect the equation of motion for $\theta$,  
which becomes 
\begin{equation}
 \frac{\diff^2 \bar{\theta} (x)}{\diff x^2} = 0 ,
\end{equation}
yielding the following solution:
\begin{equation}
 \bar{\theta} (x) = c x + d
 \with
 c, d \in \Rbb,
\end{equation}
where $c$ and $d$ are integration constants.
Although all these solutions with arbitrary values of 
$c, d$ are solutions of the field equations, they can 
give different energy density because of the total derivative term induced by the DM interaction.
We can minimize the energy density of these solutions
\begin{equation}
 {\cal E}[\bar \theta]=
 f_{\rm s}^2 \left( \frac{c^2}{2} 
\pm \kappa c \right) 
, 
 \label{eq:energyDensity-spiral}
\end{equation}
to find the ground state at the value of 
$c=\mp \kappa$. 
Since both signs give physically equivalent ground state, we 
find the spiral ground state with a moduli parameter $d$ as
\begin{equation}
\bar \phi=\frac{\pi}{2}, \quad 
\bar{\theta} (x) = - \kappa x + d, 
  \with 
  d \in \Rbb.
\label{eq:spiral-sol}
\end{equation}
Since this ground state solution represents 
the one-dimensional spiral modulation of the spin vector, 
it is called the spiral phase (see Fig.~\ref{fig:spiral-config}). 
The most general spiral solution can be obtained by applying 
simultaneous rotation in the $x$-$y$ plane and spin vector 
in the $n^1$-$n^2$ plane. 
The spiral state is similar to the helical state 
in that both describe the one-dimensional modulations. 
Nevertheless, the behavior of the collective excitation, 
or the translational phonon, is qualitatively different, 
as will be shown below. 
As a representative spiral state, we take the solution in 
Eq.~\eqref{eq:spiral-sol} as the background to study 
the dispersion relation of low-energy excitations. 

To investigate the low-energy excitation in the spiral ground state, 
we introduce the fluctuations $\delta \theta$ and 
$\delta \phi$ around the helical ground state as 
\begin{equation}
 \theta (t,\bx) = - \kappa x + \delta \theta (t,\bx) , \quad 
  \phi (t,\bx) = \frac{\pi}{2} + \delta \phi (t,\bx). 
\end{equation}
Then, we substitute this parametrization to the effective Lagrangian 
\eqref{eq:Lagrangian-isotropic} and collect the terms within 
the quadratic order of fluctuations. 
We find that it is useful to use 
$\delta \Omega (x) \equiv \sin (-\kappa x) \delta \phi$ instead 
of $\delta \phi(x)$. 
The resulting effective Lagrangian for the fluctuations
$\delta \theta$ and $\delta \Omega$ is given by
\begin{equation}
 \begin{split}
  \Lcal_{\eff}^{(2)}
  =& \frac{f_{\rm t}^2}{2} 
  \big[ 
  (\partial_0 \delta \theta)^2 + (\partial_0 \delta \Omega)^2
  \big]
  - \frac{f_{\rm s}^2}{2} 
  \big[ 
  (\partial_i \delta \theta)^2
  + (\partial_i  \delta \Omega)^2 
  \big] 
  - \frac{f_{\rm s}^2\kappa^2}{2} (\delta \Omega)^2
  \\
  & + m \delta \theta \partial_0 \delta \Omega
  - 2 f_{\rm s}^2\kappa \sin \kappa x \, \delta \theta \partial_y \delta \Omega.
 \end{split}
 \label{eq:Lagrangian-spiral-fluctuation}
\end{equation}
This result shows that the fluctuations 
$\delta \theta$ and $\delta \Omega$ 
couple through the first-order time derivative term, and the 
sinusoidal modulation proportional to the magnitude of the DM 
interaction $\kappa$ with the momentum $\partial_y$ perpendicular to the 
modulation. 
Due to the explicit presence of the sinusoidal function, the 
linear mode analysis will be a little complicated in the same 
way as the band theory with a periodic potential.

Let us investigate the low-energy spectrum described by 
Eq.~\eqref{eq:Lagrangian-spiral-fluctuation}. 
First of all, we derive the equation of motion given by 
\begin{equation}
 \begin{split}
  f_{\rm t}^2 \partial_0^2 \delta \theta 
  - m \partial_0 \delta \Omega
  - f_{\rm s}^2\bnab^2 \delta \theta 
  + 2 f_{\rm s}^2\kappa \sin \kappa x \partial_y \delta \Omega 
  &= 0,
  \\
  f_{\rm t}^2 \partial_0^2 \delta \Omega
  + m \partial_0 \delta \Omega 
  - f_{\rm s}^2\bnab^2 \delta \Omega 
  + f_{\rm s}^2 \kappa^2 \delta \Omega 
  - 2 f_{\rm s}^2\kappa \sin \kappa x \partial_y \delta \theta 
  &= 0.
 \end{split}
\end{equation}
Performing the Fourier transformation with respect to the time 
argument, we can rewrite these equations in a matrix form:
\begin{equation}
 A (\omega) \vec{\varphi}
 = H (x) \vec{\varphi}
 \with
 \vec{\varphi} \equiv 
 \begin{pmatrix}
  \delta \theta \\
  \delta \Omega
 \end{pmatrix} ,
 \label{eq:reduced-eigenproblem}
\end{equation}
where we introduced the coefficient matrices as
\begin{equation}
 A (\omega) \equiv 
  \begin{pmatrix}
   f_{\rm t}^2 \omega^2 & - \rmi m \omega  \\
   \rmi m \omega & f_{\rm t}^2 \omega^2 
  \end{pmatrix}
  \quad \mathrm{and} \quad
  H (x) \equiv f_{\rm s}^2
  \begin{pmatrix}
   - \bnab^2 & 2 \kappa \sin \kappa x \partial_y \\
   - 2 \kappa \sin \kappa x \partial_y & - \bnab^2 + \kappa^2
  \end{pmatrix}.
  \label{eq:reduced-matrix}
\end{equation}

Let us first derive the eigenvalue spectra of the reduced Hamiltonian 
$H(x)$, which are periodic along the $x$-direction as  $x \to x + a$ 
with the period $a \equiv 2\pi /\kappa$. 
Thanks to the periodicity, we can apply the Bloch's theorem~\cite{Bloch1929}, 
by introducing $\vec{\varphi}_k$ 
as a simultaneous eigenstate for $H$ and the discrete translation 
$T_a = \rme^{a \partial_x }$ as 
\begin{equation}
 H (x) \vec{\varphi}_{k_{x}} (x) 
  = E_{k_{x}} \vec{\varphi}_{k_{x}} (x)
  \quad \mathrm{and} \quad 
  T_a \vec{\varphi}_{k_{x}} (x) = \rme^{\rmi k_{x} a}
  \vec{\varphi}_{k_{x}} (x).
\end{equation}
Here, the discrete translation operator induces 
$T_a H (x+a) T_a^{-1} = H(x)$ and 
$T_a \vec{\varphi} (x) = \vec{\varphi} (x+a)$.
The Bloch's theorem tells us that we can decompose such an eigenvector as 
\begin{equation}
 \vec{\varphi}_{k_x} (x)
  = \int \frac{\diff k_{\perp}}{2\pi}
  \sum_n \rme^{\rmi (k_x + \kappa n) x + \rmi k_{\perp} y } 
  \vec{v}_{n} (\bk),
\end{equation}
with $\vec{v}_{n} (\bk) \equiv (v_{n}^{(0)} (\bk) , v_{n}^{(1)} (\bk) )^t$.
We also have introduced the momentum perpendicular to the modulation 
direction as $k_\perp$.
Note that the momentum along the modulation direction $k_x$ 
takes a value within the first Brillouin zone: 
$k_x \in [ - \pi/a, \pi/a ) = [ - \kappa/2, \kappa/2 )$.
Substituting this vector into the eigenvalue problem, 
we obtain recurrence relations among $v_n$ as
\begin{equation}
 \begin{split}
  f_{\rm s}^2 \left(
  [ (k_x +\kappa n)^2 + k_{\perp}^2 ] v_{n}^{(0)} (\bk) 
  + \kappa k_{\perp} [ v_{n-1}^{(1)} (\bk) - v_{n+1}^{(1)} (\bk) ]\right) 
  &= E_n (\bk)
  v_{n}^{(0)} (\bk) , 
  \\ 
  f_{\rm s}^2 \left(
  - \kappa k_{\perp} [ v_{n-1}^{(0)} (\bk) - v_{n+1}^{(0)} (\bk) ] 
   + [ (k_x+\kappa n)^2 + k_{\perp}^2 + \kappa^2 ] v_{n}^{(1)} (\bk) \right) 
  &= E_n (\bk)
   v_{n}^{(1)} (\bk) .
 \end{split}
\end{equation}
As is expected, the nondiagonal element is proportional to $\kappa k_\perp$.
Thus, we can derive the exact result for the eigenvalue with 
the eigenfunction on the momentum plane $k_\perp = 0$ as 
\begin{equation}
 E_n^{(0)} (k_x,0) =  f_{\rm s}^2( k_x + \kappa n )^2 
  \quad \mathrm{and} \quad 
  E_n^{(1)} (k_x,0) =  f_{\rm s}^2[( k_x + \kappa n )^2 + \kappa^2] .
\label{eq:decoupled_spectra}
\end{equation}
It is clear that the former branch of the solution with $n=0$ 
gives the lowest-lying eigenvalue, and all the bands with $n \neq 0$ 
have the gaps determined by the magnitude of the DM interaction $\kappa$.

\begin{figure}[b]
 \centering
 \includegraphics[width=0.4\linewidth]{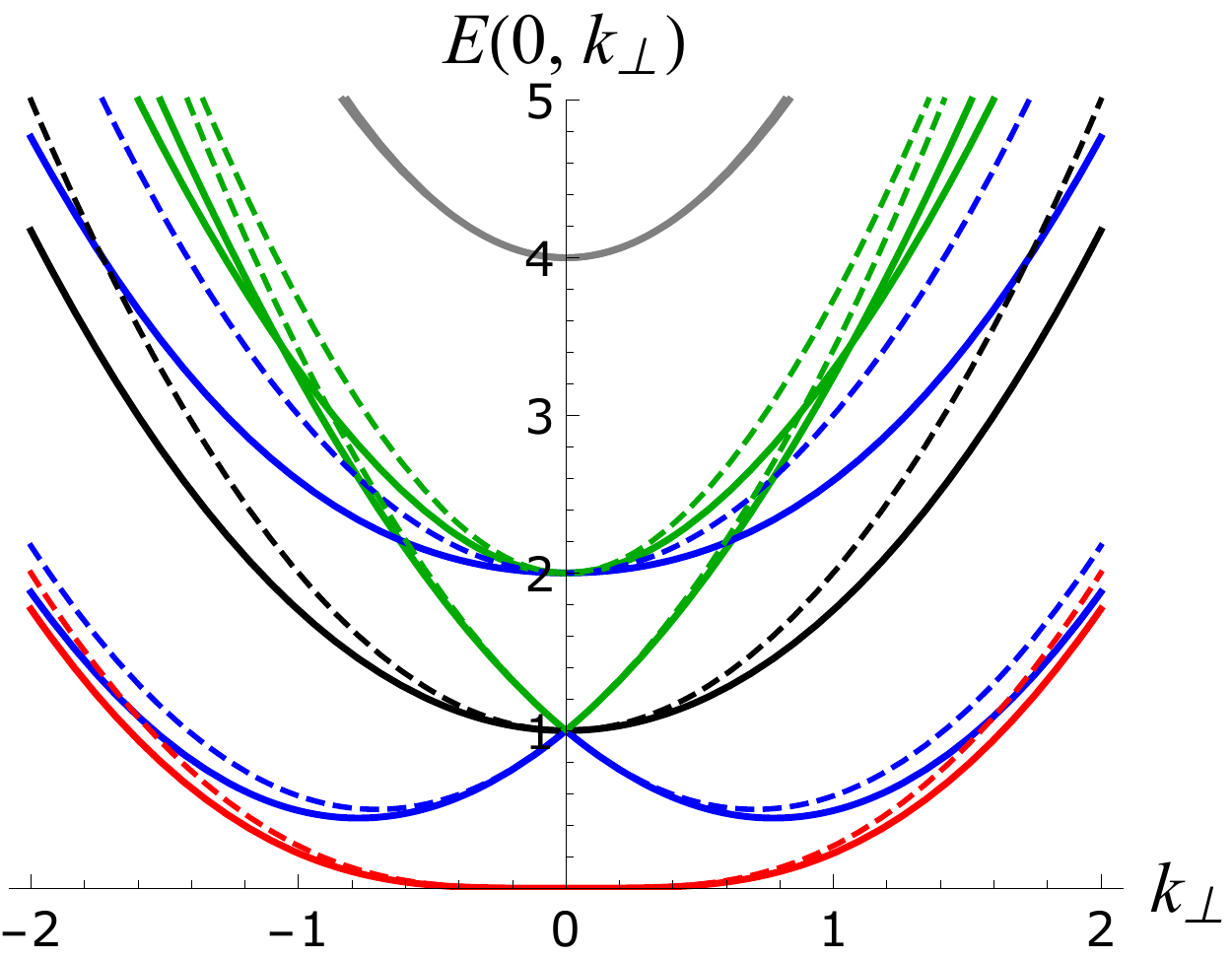}
 \caption{
 A comparison of the eigenvalue $E (0,k_\perp)$ with the $3$-band 
approximation (dashed lines) vs $5$-band approximation 
(solid lines) on the $k_x=0$ plane ($\kappa=1$).} 
\label{fig:band-3vs5}
\end{figure}

Apart from the $k_\perp = 0$ plane, we need to solve the coupled 
infinite-dimensional recurrence relation.
We observe that the coupling between neighboring bands 
$n$ and $n+1$ is proportional to $\kappa k_{\perp}$, and that 
the recurrence relations separate into two sets: those relating 
$v_{2n}^{(0)}$ with $v_{2n+1}^{(1)}$ and those relating 
$v_{2n}^{(1)}$ with $v_{2n+1}^{(0)}$. 
These facts allow us to use an  approximation to take account 
of only $2n+1$ bands between the $-n$-th and $n$-th bands, in order 
to obtain eigenvalues of the Hamiltonian at small 
$\bk^2/\kappa^2$ for low-lying states. 
Defining 
$\omega^+_n \equiv (k_x + n \kappa )^2 + k_\perp^2 + \kappa^2 $ and 
$\omega^-_n \equiv (k_x + n \kappa )^2 + k_\perp^2  $, 
we find an explicit form of the eigenvalue problem in the 
three-band truncated approximation as the following 
two sets of $3 \times 3$ matrix equations 
\begin{equation}
   \begin{pmatrix}
  \omega_1^+ - \frac{E (\bk)}{f_{\rm s}^2} & - \kappa k_\perp & 0  \\
- \kappa k_\perp & \omega_0^- - \frac{E (\bk)}{f_{\rm s}^2} & \kappa k_\perp \\
  0 & \kappa k_\perp & \omega_{-1}^+ - \frac{E (\bk)}{f_{\rm s}^2}
  \\
  \end{pmatrix}
  \begin{pmatrix}
   v_{1}^{(1)} (\bk) 
   \\
   v_{0}^{(0)} (\bk) 
   \\ 
   v_{-1}^{(1)} (\bk) 
   \\
  \end{pmatrix}
  = 0,
\end{equation}
\begin{equation}
   \begin{pmatrix}
 \omega_1^- - \frac{E (\bk)}{f_{\rm s}^2} & \kappa k_\perp & 0  \\
\kappa k_\perp & \omega_0^+ - \frac{E (\bk)}{f_{\rm s}^2} & - \kappa k_\perp \\
    0 & - \kappa k_\perp  & \omega_{-1}^- - \frac{E (\bk)}{f_{\rm s}^2} 
   \\
  \end{pmatrix}
  \begin{pmatrix}
   v_{1}^{(0)} (\bk) 
   \\ 
   v_{0}^{(1)} (\bk) 
   \\
   v_{-1}^{(0)} (\bk) 
   \\
  \end{pmatrix}
  = 0.
\end{equation}
We find discrete energy bands $E_n(\bk)$ labeled by $n=0, 1,2,\cdots$, 
as a function of momentum $\bk$ in the first Brillouin zone, by 
solving the third-order equations for vanishing determinant of 
the three-band equations.  
Similarly, we can also consider five-band truncated approximation. 
The ground state eigenvalue in the five-band approximation is 
obtained by solving the following $5\times 5$ matrix equations 
\begin{equation}
 \begin{pmatrix}
  \omega_2^+ - \frac{E (\bk)}{f_{\rm s}^2} & - \kappa k_\perp & 0  & 0  & 0  \\
  \kappa k_\perp &  \omega_1^+ - \frac{E (\bk)}{f_{\rm s}^2} & - \kappa k_\perp 
  & 0  & 0  \\
  0 & - \kappa k_\perp & \omega_0^- - \frac{E (\bk)}{f_{\rm s}^2} 
  & \kappa k_\perp & 0 \\
  0 & 0 & \kappa k_\perp & \omega_{-1}^+ - \frac{E (\bk)}{f_{\rm s}^2}
  & - \kappa k_\perp \\
  0 &0 & 0 & \kappa k_\perp & \omega_{-2}^+ - \frac{E (\bk)}{f_{\rm s}^2}
  \\
 \end{pmatrix}
  \begin{pmatrix}
   v_{2}^{(0)} (\bk) 
   \\
   v_{1}^{(1)} (\bk) 
   \\
   v_{0}^{(0)} (\bk) 
   \\ 
   v_{-1}^{(1)} (\bk) 
   \\
   v_{-2}^{(0)} (\bk) 
   \\
  \end{pmatrix}
  = 0.
\label{eq:5band_approx}
\end{equation}
Figure \ref{fig:band-3vs5} shows the comparison of the eigenvalues 
$E_n (0,k_\perp)$ with the three-band and five-band approximations. 
Note that while the results for the three-band approximation (dashed lines) 
and five-band approximation (solid lines) are not so different at 
the low-$k_\perp$ region and the low-lying band, the deviation appears 
at high-$k_\perp$ regions and at higher bands. 
As we increase the number of bands in the approximation, we, of course, obtain more bands of eigenvalues. 
We are interested in the dispersion relation at small values of 
momentum, especially for low-lying states. 
Because the coupling between neighboring bands is proportional to 
$k_\perp/\kappa$, we can use an expansion in powers of $k_\perp/\kappa$ 
to evaluate energy eigenvalues. 
We find that the $(2n+1)$-band approximation gives an exact 
result for the lowest-band spectrum $E_0(\bk)/(f_{\rm s}^2\kappa^2)$ 
up to terms of order $(k_\perp/\kappa)^{2n}$ in powers of 
$(k_\perp/\kappa)^2$.

Once eigenvalues of the reduced Hamiltonian $H$ are given in 
terms of the band energy spectra $E_n(\bk)$ in the momentum space 
with the corresponding eigenvector $\vec{\varphi}_n$, 
we can obtain the dispersion relation by solving the following equation [recall Eqs.~\eqref{eq:reduced-eigenproblem}-\eqref{eq:reduced-matrix}]:
\begin{equation}
 \begin{pmatrix}
  f_{\rm t}^2 \omega^2 - E_n (\bk) & - \rmi m \omega  \\
  \rmi m \omega & f_{\rm t}^2 \omega^2 - E_n (\bk)
 \end{pmatrix}
 \vec{\varphi}_n = 0.
\end{equation}
To find nontrivial eigenvectors, we require the 
determinant of the coefficient matrix to vanish. 
This characteristic equation gives the dispersion relations given by
\begin{equation}
 \begin{split}
  &\bullet~\mbox{Antiferromagnet}~
  (f_{\rm t} \neq 0,~m = 0):~
  \omega_n (\bk) 
= \frac{\sqrt{E_n (\bk)}}{f_{\rm t}} 
=c_s \frac{\sqrt{E_n (\bk)}}{f_{\rm s}} 
  \quad (2~\mathrm{modes}) ,
  \vspace{5pt} \\
  &\bullet~\mbox{Ferromagnet}~
  (f_{\rm t} = 0,~m \neq 0):~
  \hspace{7mm} \omega_n (\bk) = 
\dfrac{E_n (\bk)}{m} 
  \quad (1~\mathrm{mode}) ,
  \vspace{5pt} \\
  &\bullet~\mbox{Ferrimagnet}~
  (f_{\rm t} \neq 0,~ m \neq 0):~
 \hspace{8mm} \omega_n(\bk) =
\dfrac{\sqrt{m^2 + 4 f_{\rm t}^2 E_n (\bk)}\pm m}
{2f_{\rm t}^2}
  \quad (2~\mathrm{modes}) ,
 \end{split}
 \label{eq:band-3cases}
\end{equation}
where we again introduced $c_s \equiv f_{\rm s}/f_{\rm t}$.
As is the case for the homogeneous ground state, 
the number of the independent modes for ferromagnets 
($f_{\rm t} \neq 0,~ m = 0$) is half of that of the antiferromagnets 
($f_{\rm t} = 0,~ m \neq 0$) or ferrimagnets 
($f_{\rm t} \neq 0,~ m \neq 0$). 
This is because the vanishing $f_{\rm t}^2$ makes two fluctuation 
components $\delta \theta $ and $\delta \Omega$ to be one 
canonically conjugate pair of dynamical variables so that 
they just give one independent degree of freedom in contrast to 
the case of other magnets, where they become two independent degrees.

\begin{figure}[t]
 \centering
 \includegraphics[width=1.0\linewidth]{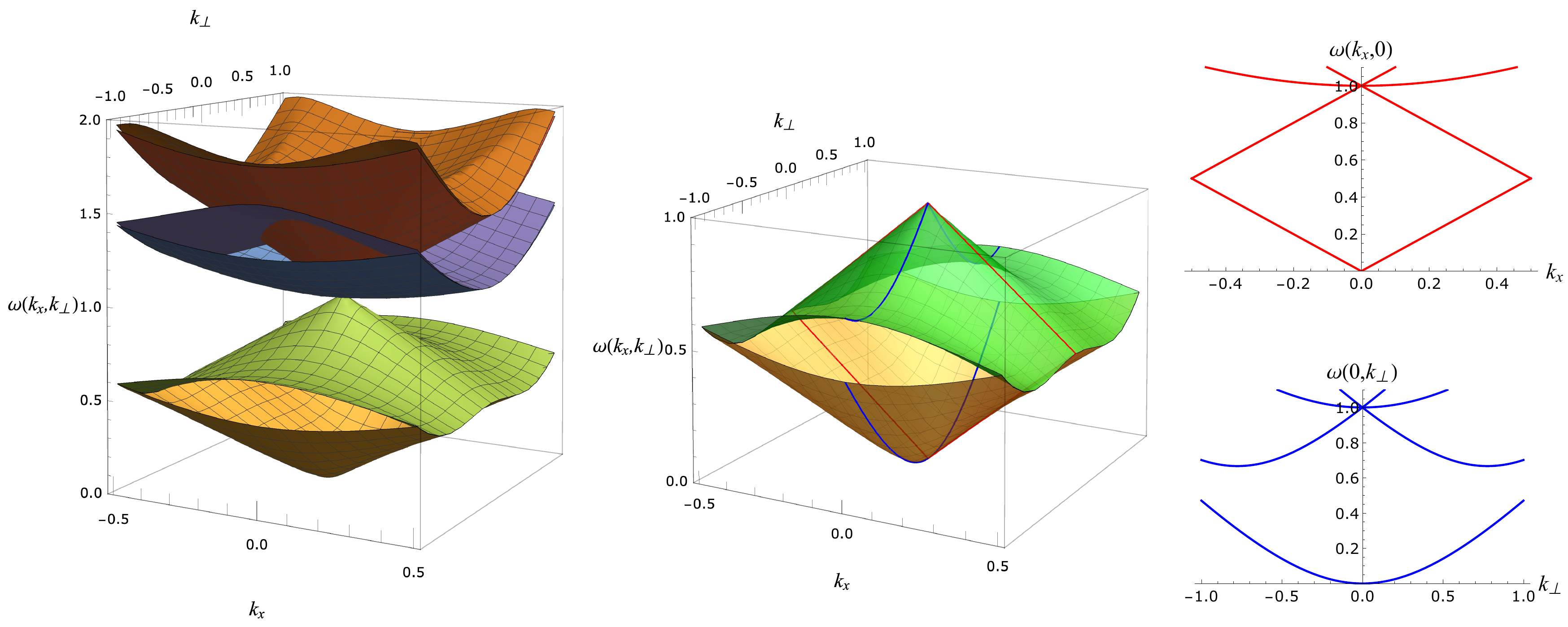}
 \caption{
 Low-energy spectrum for the antiferromagnetic spiral phase with the $5$-band approximation ($\kappa=1,~f=1$).
 } 
\label{fig:antiferro-spiral}
\end{figure}

\begin{figure}[htb]
 \centering
 \includegraphics[width=1.0\linewidth]{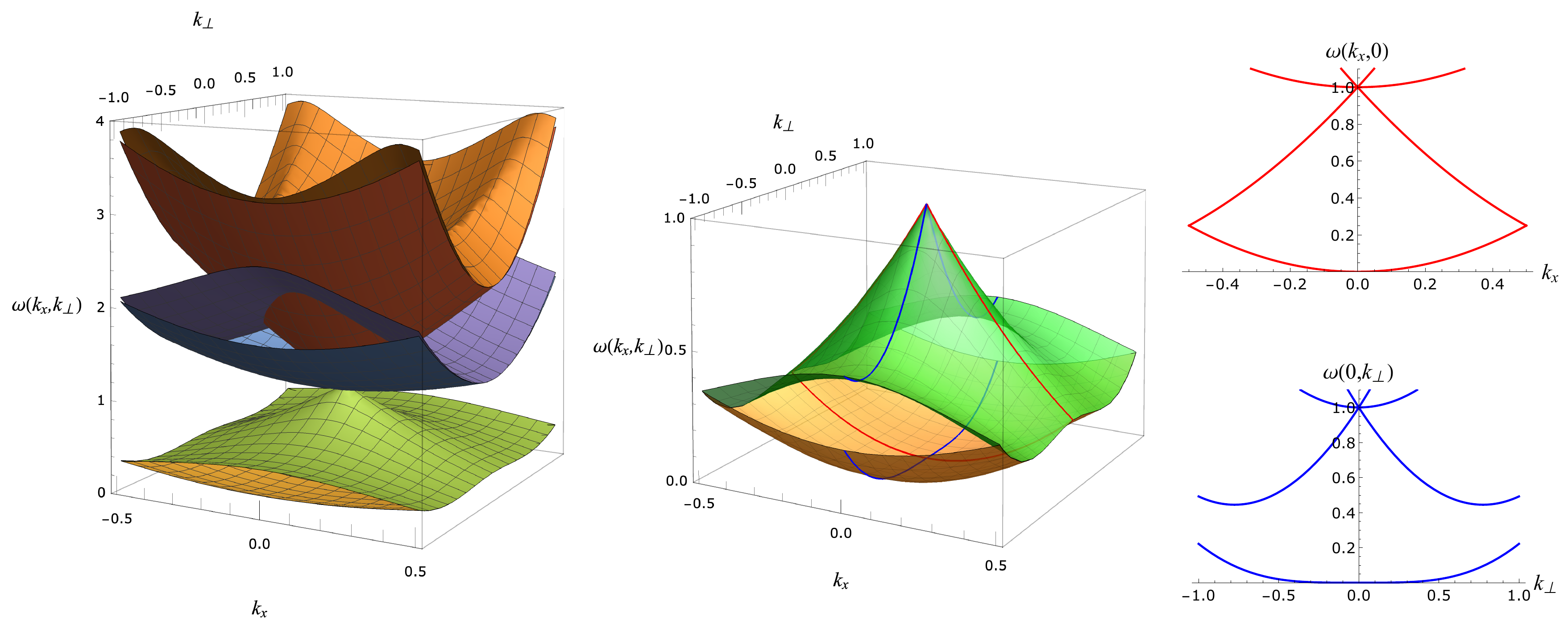}
 \caption{
 Low-energy spectrum for the ferromagnetic spiral phase with the $5$-band approximation ($\kappa=1,~m=1$).
 } 
\label{fig:ferro-spiral}
\end{figure}

\begin{figure}[htb]
 \centering
 \includegraphics[width=1.0\linewidth]{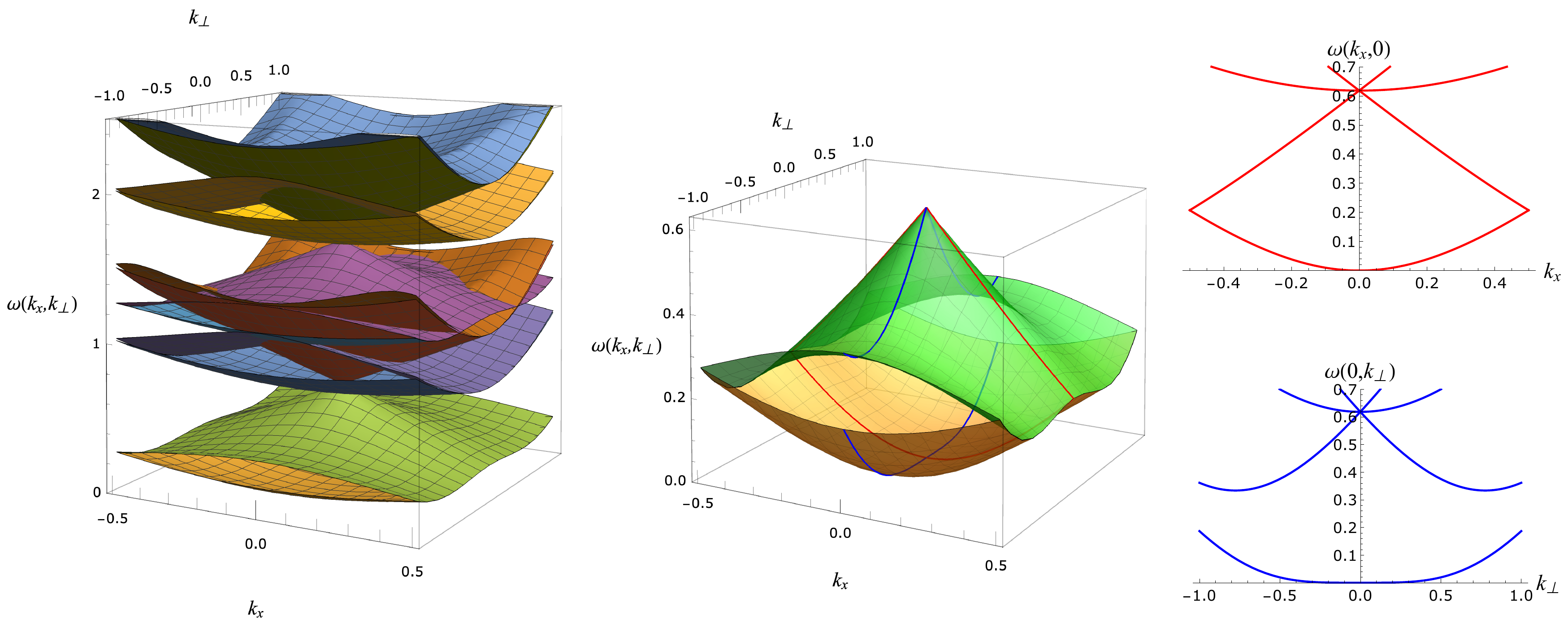}
 \caption{
 Low-energy spectrum for the ferrimagnetic spiral phase with the $5$-band approximation ($\kappa=1,~f^2=1,~m = 0.5$).
 } 
\label{fig:ferri-spiral1}
\end{figure}

\begin{figure}[htb]
 \centering
 \includegraphics[width=1.0\linewidth]{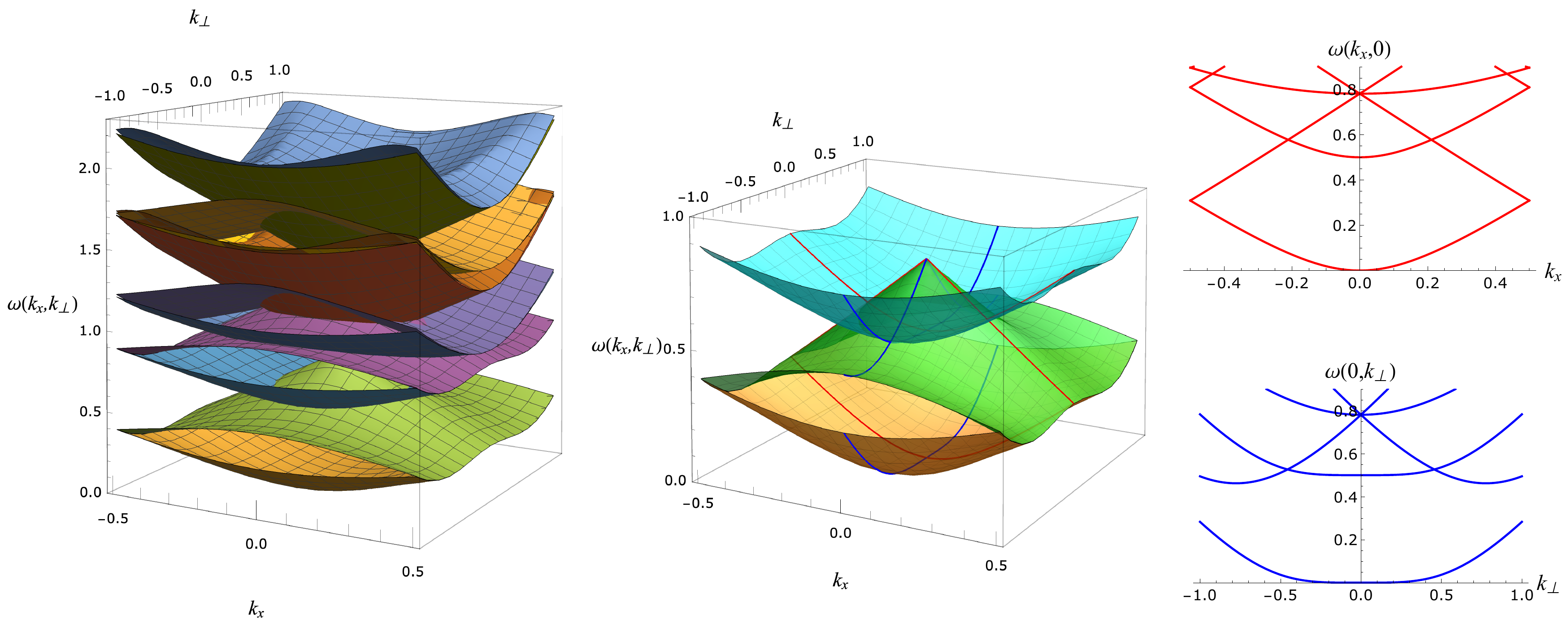}
 \caption{
 Low-energy spectrum for the ferrimagnetic spiral phase with the $5$-band approximation ($\kappa=1,~f^2=1,~m=1$).
 } 
\label{fig:ferri-spiral2}
\end{figure}

Equation~\eqref{eq:band-3cases} enables us to clarify the low-energy 
spectrum of the spiral phase from the approximated eigenvalue 
$E_n (k_x,k_\perp)$.
The resulting dispersion relations with the five-band approximation
are shown in Figs.~\ref{fig:antiferro-spiral}-\ref{fig:ferri-spiral2} 
(note that $k_x$ takes the value in 
$k_x \in [ - \kappa/2, \kappa/2 )$ 
while $k_\perp$ can be any real number $k_\perp \in \Rbb$).
One sees that the lowest branch of the bands ($n=0$) 
gives the gapless excitation, 
which dominates the low-energy behavior of the spiral phase. 
Besides, we also have other bands ($n= \pm 1, \pm2$ in the 
current working accuracy) corresponding 
to the gapped excitation 
in the first Brillouin zone: $k_x \in [ - \kappa/2, \kappa/2 )$.
Recall that the number of the independent mode is different as 
shown in  Eq.~\eqref{eq:band-3cases}: all the spectra for the 
antiferromagnetic case ($f_{\rm t} \neq 0,~ m = 0 $) are doubly 
degenerated, and the ferrimagnetic case ($f_{\rm t} \neq 0,~ m \neq 0 $) 
breaks that degeneracy, so that  
more surfaces can be seen 
in the leftmost panels of Figs.~\ref{fig:ferri-spiral1}-\ref{fig:ferri-spiral2}.

The rightmost panel in 
Figs~\ref{fig:antiferro-spiral}-\ref{fig:ferri-spiral2} shows 
the section of the low-energy 
spectrum at $k_x=0$ and $k_\perp = 0$, respectively.
In sharp contrast to the helical phase, the fluctuation spectrum 
at the low-energy region shows anisotropic behaviors.
This peculiar behavior results from the anisotropic behavior 
of the eigenvalue $E_{n=0} (k_x,k_\perp)$:
\begin{equation}
 E_{n=0} (k_x, k_\perp) = k_x^2 
-\frac{k_x^2 k_\perp^2}{\kappa^2}
+ \frac{3k_\perp^4}{8\kappa^2}  + \cdots,
\end{equation}
which is exact up to the order of $k_\perp^4$ in powers of 
$k_\perp/\kappa$, and can be obtained in the five-band approximation 
in Eq.~\eqref{eq:5band_approx}. 

Using Eq.~\eqref{eq:band-3cases}, we find the low-energy spectrum 
for the spiral phase. 
Depending on the type of magnets, we find the dispersion relation 
for the lowest (gapless) mode as follows
\begin{itemize}
 \item
 Antiferromagnet:
 \begin{equation}
  \omega_{n=0}
   (\bk) = 
   \begin{cases}
    c_s |k_x|
    \left( 1- \dfrac{k_\perp^2}{2\kappa^2} 
    + \dfrac{3k_\perp^4}{16\kappa^2|k_x|^2} + \cdots 
    \right)
    \quad \mathrm{if} \; k_x\not=0, \; \; 
    \vspace{3pt} \\
    c_s \sqrt{\dfrac{3}{8}} \dfrac{k_\perp^2}{\kappa} + \cdots
    \quad \mathrm{if} \; k_x=0, 
   \end{cases} 
 \end{equation}
 \item
      Ferromagnets or Ferrimagnets:
      \begin{equation}
 \omega_{n=0}
 (\bk) = 
\dfrac{f_{\rm s}^2}{m}\left[
 k_x^2\left(1-\frac{k_\perp^2}{\kappa^2}\right) 
+ \frac{k_\perp^4}{2\kappa^2} +\cdots \right] .
      \end{equation}
\end{itemize}
This gapless excitation is identified as the NG mode associated 
with the spontaneous symmetry breaking of the 
one-dimensional translation.
The anisotropic dispersion relation is a remarkable property of 
the one-dimensional modulation consistent with the result from a 
symmetry-based general approach~\cite{Hidaka:2014fra}. 
We also note that ferrimagnets have another branch of the gapped 
mode, whose gap is controlled by the magnetization parameter $m$. 
Thus, the gapped mode can appear with a relatively small gap 
when $m/(f_{\rm s}f_{\rm t}\kappa)<1$ (compare 
Fig.~\ref{fig:ferri-spiral1} and Fig.~\ref{fig:ferri-spiral2}).

A remark on the possible breakdown of the long-range order is in order.
Due to its peculiar low-$k_\perp$ behavior of the dispersion relation---quadratic for antiferromagnet and quartic for ferro/ferrimagnets---one may wonder whether it does affect the fate of the spiral phase to be disordered or to be the quasi-long range order.
At zero temperature, we may not encounter with an infrared divergence for the correlation function of NG modes thanks to the frequency (and $k_x$) integral.
In particular, the zero-temperature ferro/ferrimagnets are free from such a dangerous fluctuation contribution because one finds no contribution after performing frequency integral.
This situation is similar to the fact that the Mermin-Wagner theorem~\cite{Mermin:1966fe,Hohenberg:1967zz,Coleman:1973ci} does not apply to the zero-temperature $(1+1)$-dimensional ferromagnet.
Nevertheless, in the finite-temperature systems, all magnets could suffer from the divergent fluctuation contribution, so that they may develop the quasi-long range order (or may be disordered) instead of the true long-range order~(see e.g.,~\cite{Chaikin2000}).
It is interesting to investigate the fate of the spiral phase at the finite-temperature, but it is beyond the scope of this paper%
\footnote{
See Ref.~\cite{Radzihovsky:2011rf} for a recent discussion for the fate of the Fulde-Ferrell-Larkin-Ovchinnikov superfluid phase.
}.

Before closing this section, we comment on the degenerate point 
in the spectrum.
As we learn at the beginning of the band theory, the band crossing 
is usually avoided because of the level repulsion, which results 
from nondiagonal matrix elements of the involving energy states.
However, as is shown in Figs.~\ref{fig:antiferro-spiral}-\ref{fig:ferri-spiral2}, we find several crossing points at the higher bands.
This is because every band in the current model only couples to 
their nearest neighboring bands so that no level repulsion takes 
place between non-nearest neighbors.
In that sense, most of the degenerate points appear just accidentally.
However, there is a special degenerate point 
$(k_x,k_\perp) = (\pm \kappa/2,0)$ located at the boundary of the 
Brillouin zone, where all bands show two-fold degeneracy.
This two-fold degeneracy has a simple origin: 
first, there is no coupling between different bands when 
$k_\perp = 0$, so that the $k_x$ spectrum on the $k_\perp = 0$ 
plane is continuous as given in Eq.~\eqref{eq:decoupled_spectra}.
Second, the $k_x$ spectrum has to live within the first Brillouin 
zone, because of the periodicity in the $x$ direction. 
Therefore, the two-fold degeneracy at $(k_x,k_\perp) = (\pm \kappa/2,0)$ 
is inevitable.

\section{Magnon production by inhomogeneous magnetic field}
\label{sec:Pair}

In this section, 
we consider the production rate of magnons 
from the homogeneous ground state in $(d+1)$-dimensions 
caused by an inhomogeneous magnetic field, 
as another application of 
the effective field theory of magnons. 
This mechanism gives a magnon analogue 
of a pair creation of charged particles 
by an electric field---the Schwinger mechanism \cite{Schwinger:1951nm}. 
We show that the magnon production rate 
(or ground state decay rate) 
is controlled by an ``effective mass'' of the magnon consisting 
of the quadratic term of the potential and 
the ratio of the coefficients of the linear and quadratic time derivative terms. 
Hence, we will find different types of magnets 
(ferro-, antiferro-, ferrimagnets) give 
drastically different magnon production rates. 

Suppose that our spin system possesses a potential with an easy-axis 
anisotropy and develops the homogeneous ground state.
In addition, we apply an inhomogeneous magnetic field along 
the spin direction of the ground state 
and investigate the resulting dynamics of magnons. 
We also assume for simplicity that there is no DM interaction. 
This situation is described by the effective Lagrangian with 
the following background values of the external fields:
\begin{equation}
 A_0^a = \mu B (x) \delta^a_3 , \quad A_i^a = 0, 
  \quad 
  \mathrm{and} \quad 
  \ell W^{ab} = \frac{M^2}{2} \delta^a_3 \delta^b_3 ,
\label{eq:backgr_magnonProd}
\end{equation}
with $M^2> 0$ (easy-axis).  
We also assume the sign of the background magnetic field 
as $B(x) \ge 0$, so that the ground state is fixed as $n^a=(0,0,1)^t$.
Substituting these background values
into the effective Lagrangian \eqref{eq:EffLag1}, 
we obtain the effective Lagrangian 
at the quadratic order of fluctuation fields 
$\pi^\alpha ~(\alpha=1,2)$ around 
the ground state $\average{n^a}=(0,0,1)^t$ as 
\begin{equation}
 \begin{split}
  \Lcal^{(2)}
  &= - \frac{m}{2} \epsilon^3_{~\alpha\beta} \pi^\alpha \partial_0 \pi^\beta
  + \frac{f_{\rm t}^2}{2} \delta_{\alpha\beta} D_0 \pi^\alpha D_0 \pi^\beta
  - \frac{f_{\rm s}^2}{2} \delta^{ij}\delta_{\alpha\beta} 
  \partial_i \pi^\alpha \partial_j \pi^\beta
  - \frac{m \mu B(x) + M^2}{2} (\pi^\alpha)^2 .
  \\
 \end{split}
\end{equation}

One sees that the easy-axis potential generates the mass term 
proportional to $M^2$ for the magnon.
The effect of the applied magnetic field appears inside the 
covariant time derivative and the mass term if $m \not = 0$. 
In order to obtain the production rate of magnons 
due to the inhomogeneous magnetic field, 
we only need to consider 
the above quadratic effective Lagrangian. 
Hence, we neglect the interaction term in the following.

The occurrence of the magnon pair production becomes clear 
by mapping our model to the system of 
a relativistic charged scalar field.
For that purpose, we introduce a new complex scalar field 
$\Phi$ defined by the linear combination of magnon fluctuations 
$\pi^1$ and $\pi^2$ as 
\begin{equation}
 \begin{cases}
  \Phi \equiv \dfrac{1}{\sqrt{2}} (\pi^1 + \rmi \pi^2), \\
  \Phi^* \equiv \dfrac{1}{\sqrt{2}} (\pi^1 - \rmi \pi^2),
 \end{cases}
 \eqq \quad 
 \begin{cases}
  \pi^1 \equiv \dfrac{1}{\sqrt{2}} (\Phi + \Phi^*), \\
  \pi^2 \equiv \dfrac{1}{\sqrt{2} \rmi} (\Phi - \Phi^*).
 \end{cases}
\end{equation}
This transformation enables us to rewrite the effective Lagrangian 
in terms of the complex scalar field $\Phi$ as 
\begin{equation}
 \begin{split}
  \Lcal^{(2)} =
  f_{\rm t}^2 D_0 \Phi^* D_0 \Phi
  - \frac{\rmi m}{2} 
  \left[ \Phi D_0 \Phi^*  - \Phi^* D_0 \Phi \right]
  - f_{\rm s}^2\delta^{ij} \partial_i \Phi^* \partial_j \Phi
  - M^2 \Phi^* \Phi ,
 \end{split}
 \label{eq:Complex-scalar}
\end{equation}
where the covariant derivative acting on $\pi^\alpha$ is 
translated to that acting on the complex scalar field:
\begin{equation}
 D_0 \Phi = \partial_0 \Phi + \rmi \mu B \Phi, 
  \quad 
  D_0 \Phi^* = \partial_0 \Phi^* - \rmi \mu B \Phi^*.
\end{equation}
Apart from the background scalar potential $A_0 (x) = \mu B (x)$, 
the effective Lagrangian~\eqref{eq:Complex-scalar} takes a familiar form describing a relativistic charged scalar field 
except for the linear time derivative term, 
which manifestly breaks the Lorentz symmetry 
(with an effective speed of light 
$c_s = f_{\rm s}/f_{\rm t}$). 
The present model \eqref{eq:Complex-scalar} with a general 
$f_{\rm t}^2/m$ interpolates the relativistic (quadratic time 
derivative) to a nonrelativistic (linear time derivative) 
charged scalar field~\cite{Kobayashi:2015pra}.
In fact, by changing the ratio of the low-energy coefficient 
$f_{\rm t}^2/m$, we can interpolate two limiting regimes.  
Let us denote the characteristic energy scale as $k$. 
When $f_{\rm t}^2/m \gg k$, 
one can neglect the second term, and the model~\eqref{eq:Complex-scalar} 
reduces to the usual relativistic charged scalar field. 
On the other hand, when $f_{\rm t}^2 /m \ll k$, one can instead throw away 
the first term, and the model~\eqref{eq:Complex-scalar} describes 
a bosonic Schr\"odinger field~(see, e.g., Ref.~\cite{Kobayashi:2015pra} 
for more detailed discussions).

We observe that this Lorentz-symmetry breaking term 
can be regarded as a chemical potential 
corresponding to the $\U(1) \simeq \SO(2)$ symmetry 
in the relativistic charged scalar model. 
Therefore, we can absorb the linear time derivative term 
into the temporal component of the external gauge field 
by defining a modified covariant derivative $\Dcal_\mu \Phi$ as 
\begin{equation}
 \Dcal_\mu \Phi
  \equiv \partial_\mu \Phi - \rmi \Acal_\mu \Phi , 
  \quad
  \Dcal_0 \Phi^*
  \equiv \partial_\mu \Phi^* +\rmi \Acal_\mu\Phi^*, 
 \label{eq:new_covariant_der}
\end{equation}
with an inhomogeneous scalar potential 
\begin{equation}
  \Acal_0(x) \equiv \frac{m}{2f_{\rm t}^2 } - \mu B(x) , \quad 
  \Acal_i = 0 .
 \label{eq:new_covariant_der}
\end{equation}
We can now rewrite the effective Lagrangian 
\eqref{eq:Complex-scalar} to a more useful expression 
paying a cost of a constant mass shift, 
leading to the following effective action:
\begin{equation}
 \begin{split}
  \Scal_{\eff} [\Phi;\Acal_0]
  &= \int \diff^4 x 
  \left[ f_{\rm t}^2 
  \Dcal_0 \Phi^* \Dcal_0 \Phi
  - f_{\rm s}^2 \delta^{ij} \Dcal_i \Phi \Dcal_j \Phi
  - M_{\eff}^2 \Phi^* \Phi 
  \right],
 \end{split}
 \label{eq:Action-magnetic}
\end{equation}
with the effective mass $M_{\eff}$ defined by
\begin{equation}
 M_{\eff}^2 \equiv M^2 + \dfrac{m^2}{4f_{\rm t}^2}. 
\label{eq:effective_mass}
\end{equation}
Therefore, our problem is mapped to that of 
a relativistic charged scalar field 
described by the action \eqref{eq:Action-magnetic} 
with the effective mass \eqref{eq:effective_mass} 
and the inhomogeneous external electric potential 
\eqref{eq:new_covariant_der}.

As a consequence of our mapping to the effective action 
\eqref{eq:Action-magnetic}, we can carry over all the results 
on the Schwinger mechanism for the relativistic charged scalar 
by simply replacing the electric field with
$E_i \equiv \partial_i \Acal_0 - \partial_0 \Acal_i 
= - \mu \partial_i B$ 
and the mass with $M_{\eff}^2$ (see, e.~g.,~Ref.~\cite{Gelis:2015kya} 
for a recent review on the Schwinger mechanism). 
To investigate the magnon production, 
we consider a simple inhomogeneous magnetic field profile 
linear in $x$ 
\begin{equation}
B (x) =  -b(x-x_0),
\label{eq:linear_magnetic_field}
\end{equation} 
with $b>0$. We assume that $n^3=+1$ is the homogeneous ground 
state. To assure it, we can take a finite interval of $x$,
let $x_0$ to the right of the region, and then take the limit 
of an infinitely large region ($x_0\to \infty$). 
In this limit, we obtain the positive linearly decreasing inhomogeneous 
magnetic field applied to the $n^3=+1$ homogeneous ground 
state. 
Similarly to the Schwinger mechanism of charged 
particle pair production by a constant electric field%
\footnote{In the case of charged particle production, 
the constant piece of $\Acal_0(x)$ does not affect 
the production rate because it is a gauge degree of freedom. 
However, the constant piece of $B(x)$ appearing in $\Acal_0(x)$ 
in Eq.~\eqref{eq:new_covariant_der} 
in the case of magnon production 
is physical and is used to tune 
the homogeneous ground state, although 
it does not affect the production rate. 
}, we expect to obtain the pair production rate of magnon and 
anti-magnon by this linearly decreasing magnetic field. 
We will compute this production rate in an idealized 
situation of an infinite $x$ interval in the following. 
The generating functional as a functional of the gauge potential 
$\Acal$ is given by
\begin{equation}
 Z [\Acal] = \rme^{\rmi W[\Acal]}
  \equiv \lim_{T \to \infty} 
  \bra{0} \rme^{- \rmi \hH_{\mathrm{\Phi}} (\Acal) T } \ket{0}
  = \Ncal \int \Dcal \Phi \, \rme^{\rmi \Scal_{\eff} [\Phi;\Acal]},
  \label{eq:GF}
\end{equation} 
where $\ket{0}$ denotes the vacuum state, and 
$\hH_{\mathrm{\Phi}} (\Acal)$ is the Hamiltonian of the magnon 
under the inhomogeneous magnetic field obtained from
the Lagrangian Eq.~\eqref{eq:Complex-scalar} 
($\Ncal$ is a normalization constant). 
In the language of the relativistic charged scalar field, 
we can regard the slope of the magnetic field $b$ as the applied 
constant electric field because of
$E^x = \partial_x \Acal_0 (x) - \partial_0 \Acal_x =  \mu b$. 
The generating functional \eqref{eq:GF} defines the vacuum-to-vacuum 
transition amplitude, and its imaginary part, if present, can 
be regarded as the ground state decay rate.
Thus, we will evaluate the imaginary part of the generating functional below.

One nice way to evaluate the generating functional is the worldline 
formalism, which is originally developed by 
Feynman~\cite{Feynman:1950ir,Feynman:1951gn} 
along the line of 
the proper-time formalism of the Fock and Nambu~\cite{Fock:1937dy,Nambu:1950rs} (see, e.g., Ref.~\cite{Schubert:2001he,Dunne:2005sx,Gelis:2015kya} 
for reviews on the worldline formalism). 
We use the worldline formalism, 
which will be briefly described subsequently 
in order to make the paper self-contained. 
Here, we introduce the effective Minkowski metric 
$\eta_{\mu\nu} = \diag (-f_{\rm t}^{-2}, f_{\rm s}^{-2}, \cdots,f_{\rm s}^{-2})$
and 
$\eta^{\mu\nu}=\diag (-f_{\rm t}^2, f_{\rm s}^2, \cdots, f_{\rm s}^2)$, which allows us to express the effective action in a covariant manner as 
\begin{equation}
 \begin{split}
  \Scal_{\eff} [\Phi;\Acal]
  &= \int \diff^{d+1} x 
  \Phi^* \left[ \eta^{\mu\nu} \Dcal_\mu \Dcal_\nu - M_{\eff}^2 \right] \Phi.
 \end{split}
\end{equation}
Then, performing the Gaussian integral and using $\log \det A = \Tr \log A$, 
we can rewrite the generating functional as 
\begin{equation}
 \rmi W [\Acal] = -\log 
  \det [-\Dcal^2 + M_{\eff}^2]
  =  \Tr \log [-\Dcal^2 + M_{\eff}^2],
\end{equation}
with $\Dcal^2=\eta^{\mu\nu} \Dcal_\mu \Dcal_\nu$. 
We also have introduced the normalization factor 
by putting the path-integral in the absence of 
the background field. 
Since the normalization does not play an essential role in our 
discussion, we will omit it below. 
Using a zeta function regularization, 
we obtain the following 
identity for an operator $O$ 
\begin{equation}
 \log (O-\rmi \epsilon) 
  = - \int_0^\infty \frac{\diff s}{s} 
  \rme^{\rmi s (-O
  + \rmi \epsilon)} \with 
  \epsilon > 0,
\end{equation}
where $s$ denotes the so-called proper time.
With the choice of $O = \det \frac{1}{2} [-\Dcal^2 + M_{\eff}^2]$, 
this identity enables us to express 
the generating functional in terms of the 
proper time integral as follows:
\begin{equation}
 \rmi W [\Acal] = \int_0^\infty \frac{\diff s}{s} 
  \rme^{-\epsilon s} \rme^{- \frac{\rmi}{2} M_{\eff}^2 s} \,
  \Tr \big( \rme^{-\rmi s \hH (\Acal)} \big),
\end{equation}
where we introduced the differential operator $\hH (\Acal)$ by
\begin{equation}
 \hH (A) = 
  \frac{\eta^{\mu\nu}}{2} \left[ \hp_\mu - \Acal_\mu (\hx) \right]
 \left[ \hp_\nu - \Acal_\nu (\hx) \right]
  \with
  \hp_\mu \equiv - \rmi \partial_\mu .
\end{equation} 
One can see that this differential operator is 
nothing but the Hamiltonian 
for one-particle quantum mechanics, 
where the associated degree 
of freedom is called the worldline particle.
The corresponding phase-space path-integral formula is given by 
\begin{equation}
 \begin{split}
  \rmi W[\Acal] 
  = \int_0^\infty \frac{\diff s}{s} 
  \rme^{- \epsilon s} \rme^{-\frac{\rmi}{2} M_{\eff}^2 s}
  \int \diff^{d+1} x \int_{x (0) = x(s)} \Dcal x^\mu \Dcal p_\mu
  \exp \left(
  \int_0^s \diff t \left[ \rmi p_\mu \dot{x}^\mu - \rmi H (x,p;\Acal) \right] 
  \right),
 \end{split}
 \label{eq:W-WL1}
\end{equation}
where we have imposed the boundary condition $x (0) = x (s)$ 
corresponding to the trace operation.
Equation \eqref{eq:W-WL1} gives a general path-integral formula 
for the generating functional in the worldline formalism.
The applied background field is now interpreted as the gauge 
potential acting on the worldline particle.
Thus, in the worldline formalism, the problem of evaluating the 
generating functional under the background field is translated 
into the quantum reflection problem with the corresponding potential.

In the present setup, the nonvanishing gauge field is 
$\Acal_0 = \dfrac{m}{2f_{\rm t}^2} + \mu b (x^1-x_0)$, and the other 
backgrounds are absent. 
As a result, the phase-space Lagrangian 
$L_H \equiv p_\mu \dot{x}^\mu - H (x,p;\Acal) $ is given by 
\begin{equation}
 L_H =
  p_0 \dot{x}^0
  + p_1 \dot{x}^1
  + p_{i,\perp} \dot{x}^{i,\perp}
-\frac{f_{\rm t}^2}{2} \left( p_0-\frac{m}{2f_{\rm t}^2} 
- \mu b (x^1-x_0) \right)^2
  - \frac{f_{\rm s}^2}{2} p_1^2
  - \frac{f_{\rm s}^2}{2} p_{i,\perp}^2.
\end{equation}
We can perform most of the path integral as follows.
First, the path integral with respect to the perpendicular 
variables are trivialized; namely, after performing the $x^{i,\perp}$ 
integration, we obtain $p_{i,\perp} = \const$, and performing 
the $p_{i,\perp}$ integration just shifts the normalization.
Similarly, the $x^0$-integration leads to $p_0 = \const$, but we 
keep the c-number $p_0$ integration here. 
Besides, we perform the $p_1$-integration and shift the $p_0$-integration. 
After all procedures, we eventually obtain the simplified formula 
for the generating functional
\begin{equation}
 \begin{split}
  \rmi W[A] 
  = \Ncal' L^{d-1} T \int_0^\infty \frac{\diff s}{s} 
  \rme^{- \epsilon s} \rme^{-\frac{\rmi}{2}M_{\eff}^2 s}
  \int \diff p_0 \int \diff x^1 \int_{x^1 (0) = x^1(s)} \Dcal x^1
  \exp \left(
  \rmi \Scal_{\wl} [x^1;p_0]
  \right),
 \end{split}
 \label{eq:W-WL2}
\end{equation}
where we have defined the effective action for the worldline particle as 
\begin{equation}
  \Scal_{\wl} [x^1;p_0]
   = \int_0^s \diff t 
  \left[ \frac{1}{2f_{\rm s}^2} (\dot{x}^1)^2
  - \frac{f_{\rm t}^2}{2} \left( p_0  - \mu b x^1 \right)^2
  \right].
\end{equation}
Note that the value of the worldline action associated with the 
possible classical solution, or the so-called worldline instanton, 
controls the nonperturbative contribution to the generating functional.
Thus, the remaining task is to evaluate the value of the classical 
action associated with the worldline instanton.

A direct way to evaluate the value of the classical action is 
to use the Hamilton-Jacobi equation with the help of the saddle-point approximation (recall that the solution of the Hamilton-Jacobi equation gives the value of the action).
The Hamilton-Jacobi equation in the present setup is given by
\begin{equation}
 \frac{\partial \Scal_{\wl}}{\partial s} + H_{\wl} (x^1,p_0,p_1,s) = 0 
  \quad \mathrm{and} \quad 
  p_1 = \frac{\partial \Scal_{\wl}}{\partial x^1},
\end{equation}
where $H_{\wl}$ denotes the worldline Hamiltonian defined by
\begin{equation}
 H_{\wl} = \frac{f_{\rm s}^2}{2} p_1^2 
  + \frac{f_{\rm t}^2}{2} \left( p_0  - \mu b x^1 \right)^2.
\end{equation}
Note that the worldline action enjoys the proper-time translational 
invariance, and thus, the Hamiltonian takes a constant value as
$H_{\wl} = \Ecal_{\wl} = \const$
As a result, we can solve the energy equation with respect to $p_1$ as 
\begin{equation}
 p_1(x^1) = \pm\frac{1}{f_{\rm s}} \sqrt{ 2 [ \Ecal_{\wl} - V_{\wl} (p_0;x^1)] } 
  \with 
  V_{\wl} (p_0;x^1) \equiv
  \frac{f_{\rm t}^2}{2} \left( p_0  - \mu b x^1 \right)^2.
\end{equation}
As a last step, we use the stationary phase condition for the proper time integral, which leads to $ -M_{\eff}^2/2 + \partial \Scal_{\wl}/\partial s = 0$. 
By comparing this with the Hamilton-Jacobi equation, we find the value of the energy given by $\Ecal_{\wl} = - M_{\eff}^2/2$.

Wrapping up these results, we find the solution of the Hamilton-Jacobi equation as
\begin{equation}
 \begin{split}
  \Scal_{\wl} (s) 
  &= - \Ecal_{\wl} s 
  + \frac{1}{f_{\rm s}}\oint \diff x^1 \, p_1(x^1)
  \\
  &= \frac{1}{2} M_{\eff}^2s + \frac{\rmi \pi M_{\eff}^2}
{f_{\rm s}f_{\rm t} \mu b} . 
 \end{split}
\end{equation}
where we have used $\Ecal_{\wl} = - M_{\eff}^2 $ and performed the 
contour integral to proceed to the second line.
Recalling the definition of the effective mass, we eventually 
find the leading imaginary part of the generating functional given by 
\begin{equation}
 \mathrm{Im}\, W [A] \simeq 
  \Ncal T V \exp 
  \left( 
 - \frac{\pi}{f_{\rm s}f_{\rm t} \mu b} \left( M^2 
+ \frac{m^2}{4 f_{\rm t}^2} \right) 
  \right).
  \label{eq:Decay-rate}
\end{equation}
As is expected, this agrees with the leading part of the Schwinger's 
formula for the constant electric field $E_x = \mu b$ and the 
effective mass $ M_{\eff}^2 = M^2 + \dfrac{m^2}{4 f_{\rm t}^2}$ 
with the corrections by the coefficients of time and space 
kinetic terms $f_{\rm t}$ and $f_{\rm s}$.
Note that the ratio of the low-energy coefficients $m^2/f_{\rm t}^2$ 
appears in the formula. 
As a result, the magnon production rate for antiferromagnets 
$m^2 = 0$ gives the canonical Schwinger's formula with the mass 
$M^2$ associated to the energy gap of magnon while that for 
ferromagnets $m^2/f_{\rm t}^2 \to \infty$ vanishes as 
$\mathrm{Im}\, W [A] \to 0$.
This reflects the absence of the pair production in the nonrelativistic 
systems (infinite effective mass limit).
Our result \eqref{eq:Decay-rate} for ferrimagnets with a general 
value of $m^2/f^2$ gives the interpolation between relativistic 
and nonrelativistic charged scalar fields in terms of the ground 
state decay rate.

\section{Discussion}
\label{sec:Discussion}

In this paper, we have developed a unified way to implement various $\SO(3)$ symmetry-breaking terms---the magnetic field, 
single-ion anisotropy, and DM interaction---into the low-energy 
effective field theory of spin systems.
We have also applied the constructed effective Lagrangian 
to certain situations where the symmetry-breaking terms 
induce nontrivial dynamics.
We have shown that two simple inhomogeneous ground states 
(helical and spiral phases) support the translational phonon 
as the resulting NG mode 
while they give a qualitatively different low-energy spectrum, 
such as isotropic versus anisotropic dispersion relations.
We have also discussed the analogue of 
the Schwinger mechanism 
by evaluating the decay rate of the homogeneous ground state 
induced by the inhomogeneous magnetic field.

There are several interesting prospects 
based on the present work.
One direction is to investigate various transport phenomena 
in spin systems by extending our effective Lagrangian approach.
For instance, despite the experimental realization 
of the thermal Hall effect in spin systems, 
its field-theoretical description has been still unclear.
Our formulation has a potential advantage 
to provide a direct connection between the effective field theory and the underlying lattice descriptions of spin systems.
However, it is necessary to relax our assumption 
on the cubic-type lattice 
since the thermal Hall effect takes place 
in different types of the lattice structure~(see e.g.,~\cite{Katsura2010}).
Generalization to such a nontrivial lattice may be 
important to study the thermal Hall effect 
based on the effective field theory.
Besides, it is also interesting to investigate 
the transport phenomena of spin densities, 
which lead to a potential connection to the spintronics 
(see, e.g., Ref.~\cite{Maekawa2017spin} for a review).
While the presence of small explicit breaking terms makes 
total spins as approximate conserved charges, 
its dynamics is a primary concern of the spintronics.
For example, a recent proposal in Ref.~\cite{Fujimoto2020} 
for a mechanical generation of the DM interaction 
and the resulting spin current is an interesting problem.
It is worthwhile developing the effective Lagrangian approach 
to the spintronics~(see also Refs.~\cite{Tatara2008,Tatara2019} for reviews discussing the effective Lagrangian approach to the spintronics).

Another interesting direction is to clarify a possible realization 
and resulting dynamics of the magnetic skyrmion 
based on the effective field theory%
\footnote{
The effective field theories and NG modes 
in the presence of a single magnetic skyrmion line 
\cite{Kobayashi:2014eqa}
and a single magnetic domain wall 
\cite{Kobayashi:2014xua} were discussed 
in the absence of the DM term.
}.
In $(2+1)$-dimensional cases, 
the magnetic skyrmion represents a 
nontrivial topologically stable configuration 
of the magnetization vector, 
which results in the topologically conserved charge.
As is the case for the skyrmion in hadron physics~\cite{Skyrme:1961vq,Skyrme:1962vh,Witten:1979kh}, 
it is worth understanding what conserved quantity 
this charge describes.
A natural candidate (for, at least, a particular spin systems) is 
the electric charge attached to the underlying charge carrier 
like an itinerant electron.
In such systems, when the ground state supports 
the finite local skyrmion charge 
(like the spiral phase or skyrmion crystal), 
there should be an induced electromagnetic field~\cite{Bar:2003ip,Wiese:2005tg}.
Thus, the spin could affect the dynamics of the electromagnetic field through its topological configuration, 
although it is not an electrically charged object. 
This implies a possibility of the interesting coupled dynamics of the spin and dynamical electromagnetism in a similar manner with the charge density wave phase of many-electron systems.
We leave these interesting problems as future works.

\acknowledgments

M.H. thanks Y.~Hidaka, T.M. Doi, T. Hatsuda, H. Katsura, Y. Kikuchi, K. Nishimura, Y. Tanizaki, S. Furukawa, T. Furusawa, N. Sogabe, N. Yamamoto, and H. Taya for useful discussions. 
M.H. was supported by the U.S. Department of Energy, Office of Science, Office of Nuclear Physics under Award Number DE-FG0201ER41195.
This work was supported by Japan Society of Promotion of Science (JSPS) 
Grant-in-Aid for Scientific Research (KAKENHI) Grant Numbers 18H01217, the Ministry of Education, Culture, Sports, Science, and Technology(MEXT)-Supported Program for the Strategic Research Foundation at Private Universities ``Topological Science'' (Grant No. S1511006), and the RIKEN iTHEMS Program, in particular iTHEMS STAMP working group.

\appendix

\section{Effective Lagrangian from coset construction} 
\label{sec:Coset}

In this appendix, we provide another way to construct the effective Lagrangian \eqref{eq:EffLag1}; that is, the coset construction originally developed in the context of the high-energy physics~\cite{Coleman:1969sm,Callan:1969sn,Bando:1987br}, and recently applied to magnons in Ref.~\cite{Gongyo:2016dzp}. 
We assume that the DM interaction is weaker than the potential; e.g. 
\begin{equation}
 (\kappa_i^a)^2 \ll W, B. 
  \label{eq:weakDM}
\end{equation}
This assumption allows us to start exploring the background 
(ground state) as a homogeneous state with the symmetry breaking 
pattern dictated by the potential, and to use the resulting effective 
Lagrangian to examine the effect of the DM interaction.

Suppose that the homogeneous ground state of the spin system \eqref{eq:Hamiltonian1} spontaneously breaks the approximate $\SO(3)$ symmetry down to $\SO(2)$. 
We are interested in the low-energy (long wave-length) 
behavior of the system, and we can directly employ the field-theoretical 
(continuum) description of the associated pseudo-NG mode. 
Thus, we have the background fields $A_\mu^a$ and $W$ transforming as 
the $\SO(3)$ gauge and matter field, respectively, as discussed in the main text.
The main difference is that we assume the symmetry breaking pattern, which allows us to directly introduce the NG field in the coset construction.

Let us now review how the magnon (NG field) is introduced in the effective 
Lagrangian~\cite{Gongyo:2016dzp}. 
First of all, we divide the generators of the $\so(3)$ Lie algebra 
$t_a = \{t_\alpha, t_3\}$ belonging to the broken part 
indices $\alpha = 1,2$ and unbroken $\so(2)$ index $a=3$ 
satisfying
\begin{equation}
 \tr (t_\alpha t_3) = 0, \quad  
  \tr (t_\alpha t_\beta) = g_{\alpha\beta}, \quad 
  \tr (t_3 t_3) = g_{33}, 
\end{equation}
with the Cartan metric $g_{ab}$, which reduces to $g_{ab} \to 2 \delta_{ab}$ if we choose Eq.~\eqref{eq:so(3)-generator}. 
The basic ingredient is the coset $\xi (\pi) \in \SO(3)/\SO(2)$ 
parametrizing the NG modes, or the magnons 
$\pi^\alpha$, whose representative is e.g. parametrized by
\begin{equation}
 \xi (\pi) = \rme^{\rmi \pi}, \quad 
  \pi \equiv \pi^\alpha t_\alpha. 
  \label{eq:cost_variable}
\end{equation}
using the explicit form given in Eq.~\eqref{eq:so(3)-generator}. 
We note that the local $g (x)\in \SO(3)$-transformation, by 
definition, acts on the (right-)coset element $\xi (\pi)$ as 
\begin{equation}
 \xi (\pi) \to \xi (\pi') = g (x) \xi (\pi) h^{-1} \big(\pi, g(x) \big),
  \quad
  h \big(\pi, g(x) \big) \in \SO(2).
  \label{eq:CosetTr}
\end{equation}
We then introduce the gauged Maurer-Cartan 1-form $\alpha_\mu (\pi)$ as
\begin{equation}
 \alpha_\mu (\pi) \equiv \rmi^{-1} 
  \xi^{-1} ( \pi) D_\mu \xi (\pi) 
  \with
  D_\mu \xi (\pi) \equiv \partial_\mu \xi (\pi) - \rmi A_\mu (x) \xi(\pi),
\end{equation}
with the background $\SO(3)$ gauge field 
$A_\mu = A_\mu^a t_a$, whose transformation rule is given in Eq.~\eqref{eq:BkgGaugeTr}.
With a help of Eqs.~\eqref{eq:BkgGaugeTr} and \eqref{eq:CosetTr}, we can show that the transformation rules for projected components of the Maurer-Cartan 1-form 
$\alpha_{\mu\para} \equiv \frac{1}{2} \tr (\alpha_\mu t_3) t_3 $ and 
$\alpha_{\mu\perp} \equiv \frac{1}{2} \sum_\alpha \tr (\alpha_\mu t_\alpha) t_\alpha$ 
are given by
\begin{equation}
 \begin{split}
  \alpha_{\mu\para} (\pi) 
  &\to \alpha_{\mu\para} (\pi')  
  = h \big(\pi, g(x) \big) \alpha_{\mu\para} (\pi) h^{-1} \big(\pi, g(x) \big)
  + \rmi^{-1} h \big(\pi, g(x) \big) \partial_\mu h^{-1} \big(\pi, g(x) \big),
  \\
  \alpha_{\mu\perp} (\pi) 
  &\to \alpha_{\mu\perp} (\pi')  
  = h \big(\pi, g(x) \big) \alpha_{\mu\perp} (\pi) h^{-1} \big(\pi, g(x) \big).
 \end{split}
 \label{eq:MCTr}
\end{equation}
The Maurer-Cartan $1$-form describes the NG field (magnons), which is an alternative to the normalized vector $n^a$.

We have elucidated the transformation rules of the Maurer-Cartan $1$-form and 
background fields in Eqs.~\eqref{eq:BkgGaugeTr}, \eqref{eq:CosetTr} and \eqref{eq:MCTr}.
Then, we can systematically construct the general effective Lagrangian 
once we fix the power counting scheme. 
As usual, spacetime derivatives of the NG field $\pi^\alpha(x)$ 
results in higher-order contributions to the low-energy effective 
field theory. 
We thus consider the leading-order effective Lagrangian up to 
terms with second-order derivatives of $\pi^\alpha(x)$.
This motivates us to count background fields as 
$A_\mu^a = O(\partial_\mu)$ 
and $W = O(\partial_i^2)$.
Summarizing these, we will employ the power-counting scheme:
\begin{equation}
 \pi^\alpha = O (\partial_\mu^0), \quad
  A_\mu^a = O(\partial_\mu), \quad 
  W = O(\partial_i^2), 
  \label{eq:Power}
\end{equation}
to construct the leading-order effective Lagrangian.

By the use of the above transformation rule and power-counting 
scheme, we are able to  write down the most general 
$\SO(3)$-invariant effective Lagrangian for magnons.
Here, it is important to notice that the spin system under consideration 
does not respect the Lorentz symmetry, which means that time 
and spatial components of $\alpha_{\mu\perp}$ can appear independently. 
We thus immediately find invariant terms 
$\tr (\alpha_{0\perp} \alpha_{0\perp})$ and 
$\delta^{ij} \tr (\alpha_{i\perp} \alpha_{j\perp})$ respecting 
the spatial rotation symmetry. 
Furthermore, noting that the unbroken $\SO(2)$ symmetry is abelian, 
we find an additional invariant term $\tr (t_3 \alpha_{0\para})$.
This can be explicitly shown that the general parametrization 
$h \big(\pi,g(x) \big) = \rme^{\rmi p (\pi,g(x)) t_3}$ leads 
to $h \partial_\mu h^{-1} = - \rmi t_3 \partial_\mu p (\pi,x)$, 
which means $\tr (t_3 \alpha_{0\para})$ is invariant up 
to a surface term.
Besides, a combination of the coset $\xi(x)$ and 
the background field $W(x)$ generates another invariant term.
Thanks to the relation $\tr (\xi^{-1} W \xi) = \tr W = \const$, 
we need to keep only one of two invariant terms 
$(\xi^{-1} W \xi)^{33}$ $\delta_{\alpha\beta}$ and 
$(\xi^{-1} W \xi)^{\alpha\beta}$, where indices denote that for the matrix%
\footnote{
There seems to be another invariant term $\tr (t_3 \xi^{-1} W \xi)$. 
However, this term with its complex conjugate will vanishes, and thus, 
does not appear in the effective Lagrangian.
}.
In short, we have independent four invariant terms composed of 
the gauged Maurer-Cartan 1-form:
\begin{equation}
 \tr (\alpha_{0\perp} \alpha_{0\perp}), \quad 
  \delta^{ij} \tr (\alpha_{i\perp} \alpha_{j\perp}) ,\quad
  \tr (t_3 \alpha_{0\para}), \quad
  \mathrm{and} \quad
  (\xi^{-1} W \xi)^{33}. 
  \label{eq:possible-terms-coset}
\end{equation}
Taking account of all these, we write down the general 
$\SO(3)$-invariant effective Lagrangian of magnons in the leading-order derivative expansion (up to two derivatives) as 
\begin{align}
 \Lcal_{\eff}^{(2)} 
 &= - \frac{m}{2} \tr (t_3 \alpha_{0\para}) 
 + \frac{f_{\rm t}^2}{4} \tr (\alpha_{0\perp} \alpha_{0\perp}) 
 - \frac{f_{\rm s}^2}{4} \tr (\alpha_{i\perp} \alpha_{i\perp})
 + \ell (\xi^{-1} W \xi)^{33}.
 \label{eq:EffLag-coset1}
\end{align}
Since the coset representative $\xi(\pi) = \rme^{\rmi \pi}$ 
contains an infinite number of the magnon field $\pi^\alpha (x)$, 
this effective Lagrangian describes the fully interacting model of magnons.
By expanding the coset representative $\xi(\pi) = \rme^{\rmi \pi}$ 
we obtain the effective Lagrangian \eqref{eq:EffLag1} in the main text.
One sees that four low-energy coefficients attached to four invariannts in Eq.~\eqref{eq:possible-terms-coset} indeed coincides with those appearing in the $O(3)$ nonlinear sigma model.
As discussed in the main text, their matching condition are given in Eqs.~\eqref{eq:matcthing}-\eqref{eq:matcthing2}.

\bibliography{EFT-magnon-DM.bib}

\end{document}